\documentclass[useAMS,usenatbib]{mn2e}

\title[High energy gamma-rays from Globular Clusters]
{High energy gamma-rays from Globular Clusters}
\author[W. Bednarek \& J. Sitarek]{W. Bednarek\thanks{E-mail:
bednar@fizwe4.phys.uni.lodz.pl} \& J. Sitarek\\
Department of Experimental Physics, University of \L \'od\'z,
ul. Pomorska 149/153, 90-236 \L \'od\'z, Poland}
\begin{document}

\date{Accepted . Received ; in original form }

\pagerange{\pageref{firstpage}--\pageref{lastpage}} \pubyear{2006}

\maketitle

\label{firstpage}

\begin{abstract}
It is expected that specific globular clusters can contain up to a hundred of millisecond pulsars. These pulsars can accelerate leptons at the shock waves
originated in collisions of the pulsar winds and/or inside the pulsar magnetospheres. Energetic leptons diffuse gradually through the globular cluster comptonizing stellar and  microwave background radiation. We calculate the GeV-TeV $\gamma$-ray spectra for different models of injection of leptons and parameters of the globular clusters assuming reasonable, of the order of $1\%$, efficiency of energy conversion from the pulsar winds into the relativistic leptons. It is concluded that leptons accelerated in the globular cluster cores should produce well localized $\gamma$-ray sources which are concentric with these globular clusters. The results are shown for four specific globular clusters (47 Tuc, Ter 5, M13, and M15), in which significant population of millisecond pulsars have been already discovered. We argue that the best candidates, which might be potentially detected by the present Cherenkov telescopes and the planned satellite telescopes (AGILE, GLAST), are 47 Tuc on the southern hemisphere, and M13 on the northern hemisphere. We conclude that detection (or non-detection) of GeV-TeV $\gamma$-ray emission from GCs by these instruments put important constraints on the models of acceleration of leptons by millisecond pulsars.
\end{abstract}

\section{Introduction}

Globular clusters contain $\sim$$10^5$-$10^6$ late type stars in a volume with the half-mass radius of the order of a few parsecs. They are distributed spherically around our Galaxy at characteristic distances of $\sim$10 kpc (e.g. Harris~1996). 
It is expected that the typical globular cluster can contain up to  a hundred of millisecond radio pulsars (MSP) which appear as a result of spinning up due to the accretion of matter from the low mass companion
(the so called "recycling" model, e.g. Alpar et al.~1982). In fact, more than 100 radio pulsars with spin periods in the range of a few to a few tens of milliseconds, have been detected in 24 globular clusters. The largest observed samples are from Ter 5 (23 MSP) and 47 Tuc (22 MSP) (e.g. Camilo \& Rasio~2005).
Many of these MSPs are inside the LMXB systems, but significant part is isolated probably due to the evaporation of the companion star. It is expected that MSPs produce relativistic magnetized winds, which in the case of binary pulsars interact with the stellar winds. As a result, a shock wave is created on which particles can be accelerated to energies large enough for  $\gamma$-ray production (e.g. Bigniami, Maraschi \& Treves~1977, Klu\'zniak et al.~1988, Phinney et al.~1988). Only a part of the pulsar wind is obscured due to the presence of the stellar companion.
In most directions it propagates unconfined and interacts with the globular cluster medium or with other pulsar winds. Note, that in the case of very compact binaries pulsars can be permanently
enshrouded by the evaporating material from the companion star (Tavani~1991). This class of "hidden" pulsars is difficult to detect and none of the objects of this type has been uniquely identified up to now.  

The $\gamma$-ray fluxes expected from specific binaries containing MSPs (e.g. PSR 1957+20, Arons \& Tavani~1993) or from the population of single MSPs inside the globular clusters (Tavani~1993) are about the order of magnitude below the sensitivity of the EGRET telescope. Larger fluxes are expected from the hidden MSPs in compact binaries (Tavani 1991).
In fact, only the upper limits on the  $\gamma$-ray emission from the globular clusters are available, e.g. $(1-2)\times 10^{-7}$ ph. cm$^{-2}$ s$^{-1}$ above 100 MeV 
(Michelson et al.~1994, Manandhar et al. ~1996) and $8\times 10^{-6}$ ph. cm$^{-2}$ s$^{-1}$ in energy range $0.75-30$ MeV (O'Flaherty et al.~1995). The above mentioned upper limits already allow to put constraints on the number of such sources in globular clusters.  
More recently, $\gamma$-ray emission from the inner MSP magnetospheres has been also estimated for specific GCs by Harding et al~(2005). It has been concluded that $\gamma$-rays above $50$ GeV should be observed from some specific individual MSPs and from some GCs by the MAGIC, HESS, and VERITAS Cherenkov telescopes.

The aim of this paper is to calculate the $\gamma$-ray emission in the GeV-TeV energy range expected from the whole population of MSP in the globular clusters. However, in contrast to previous models, we assume that $\gamma$-ray emission is produced by leptons
which scatter the stellar and microwave background radiation during their diffusion through
the medium of the GC. Leptons are accelerated either on the shock waves produced in collisions of the MSP winds inside the GC or injected from the inner pulsar magnetopsheres.
We show that leptons can be accelerated to high energies and are confined inside globular cluster for sufficiently long time for efficient up-scattering of the stellar and microwave background radiation to GeV-TeV energy range.
The available upper limits on the TeV $\gamma$-ray emission from the globular clusters,
e.g. in the case of M 13 (Hall et al.~2003) and M 15 (LeBohec et al.~2003), do not constrain strongly the possible production of energetic particles in these type of objects.
However, it is likely that GeV-TeV $\gamma$-rays can be detected from globular clusters by the present and future telescopes.

\section{Pulsars inside globular clusters}

The number of millisecond pulsars in the central part of typical globular cluster (GC) is expected to be of the order of a few tens up to a few hundred (e.g. Tavani 1993). In fact, 23 pulsars have been observed up to now in Ter 5 and 22 pulsars in 47 Tuc. Most of these pulsars are localized inside the core of GCs. They have the average period of the order of a few milliseconds 
and surface magnetic fields of the order of $\sim$$10^9$ G  (see e.g. Camilo \& Rasio~2005). The characteristic energy loss rate due to the pulsar wind from an isolated pulsar can be estimated by applying the formula for rotating magnetic dipole,  
\begin{eqnarray}
L_{\rm p} = 1.2\times 10^{35}B_9^2P_4^{-4}~~~{\rm erg~s^{-1}},
\label{eq1}
\end{eqnarray}
\noindent
where $B_{\rm p} = 10^9B_9$ G is the surface magnetic field of the pulsar, and $P = 4P_4$ ms is the pulsar period.

For typical radius of the core of GC, $R_{\rm c}\sim 0.5$ pc, the characteristic distances between pulsars in GC are $R_{\rm p}\approx R_{\rm c}/N_{\rm p}^{1/3}$. For example, assuming the presence of $100$  pulsars in the cluster core, we obtain $R_{\rm p}\approx 3.3\times 10^{17}$ cm. Pulsars produce relativistic strongly magnetized winds which, when colliding with other winds, create relativistic shocks. If the pulsar winds collide between themselves then typical distance from the pulsar to the shock can be estimated to be 

\begin{eqnarray}
R_{\rm sh}\sim 0.5R_{\rm p}\approx 0.5R_{\rm c}/N_{\rm p}^{1/3}. 
\label{eq1b}
\end{eqnarray}

\noindent
We estimate the magnetic field strength at such shocks by simple re-scaling from the pulsar (assuming $R^{-3}$ dependence in the inner pulsar magnetosphere and $R^{-1}$ dependence in the pulsar wind), 
\begin{eqnarray}
B_{\rm sh}\approx  3\sqrt{\sigma}B_{\rm p} (R_{\rm NS}/R_{\rm LC})^3(R_{\rm LC}/R_{\rm sh}) \nonumber \\
\approx  2.5\times 10^{-5}\sqrt{\sigma}B_9/P_4^2~~~{\rm G},
\label{eq2}
\end{eqnarray}
\noindent
where $\sigma$ is the magnetization parameter of the pulsar wind at the shock region, $R_{\rm NS} = 10^6$ cm is the neutron star radius, $R_{\rm LC} = cP/2\pi$ is the light cylinder radius. We consider the case of leptons injected by the pulsars and additionally accelerated in such relativistic shocks with the power-law spectra. 
The value of $\sigma$ has been estimated in the case of the wind from the Vela type pulsars at $\sim$0.1 (e.g. Sefako \& de Jager~2003) and from the Crab type pulsars at $\sim$0.003
(Kennel \& Coroniti~1984). If the wind is terminated closer to the pulsar, then $\sigma$ may be of the order of $\sim$1 (Contopoulos \& Kazanas~2002).
Applying this available range of $\sigma$, we estimated the magnetic field at the pulsar wind shock region in the range $B_{\rm sh}\sim 10^{-6}-10^{-5}$ G. 

The pulsar winds can also interact with the winds of the stars inside the GC. There are typically $N_{\star}\sim$2000 stars in the cores of the GCs within the spherical volume with the radius of $R_{\rm c} = 0.5$ pc. The average distance between these stars is then $R_{\star}\approx R_{\rm c}/N_{\star}^{1/3}\approx 1.2\times 10^{17}$ cm. Let us consider for simplicity solar type stars with typical mass loss rate of the order of
${\dot M} = 10^{-14}$ M$_{\odot}$ per year and the wind velocity $V_\star\sim 500$ km s$^{-1}$. We determine the parameter $\mu = L_{\rm p}/c{\dot M}V_\star$, which describes the pressure balance between the pulsar and the stellar winds. For the above values it is equal to 
$\mu\approx 1.25\times 10^5$. 
Then, based on the simplified model for the structure of the colliding winds (Girard \& Wilson~1987), we estimate the radius of the volume dominated by the stellar wind on $R_{\rm d} =  R_{\star}/(1 + \sqrt{\mu})$, i.e. equal to 
$R_{\rm d}\approx 2.8\times 10^{-3}R_{\star}\approx 3.4\times 10^{14}$ cm for a solar type star. Therefore, we conclude that stellar winds are confined in a small region around the stars and the pulsar winds collide mainly between themselves.

Globular clusters can contain typically a few hundred thousand stars with characteristic masses of the order of $\sim 1$ M$_{\odot}$ in the volume of a few parsecs. Since their mass to luminosity ratio is close to $\sim 2$ (see e.g. recent analysis for M15 by van den Bosch et al. 2006), these stars create very strong background radiation field which can act as a target for relativistic leptons injected by the millisecond pulsars.
Based on the observed luminosity of specific globular cluster, $L_{\rm GC}$, and on the density profile for the distribution of the stars inside the cluster (based on the Michie-King model (Michie 1963), see also Kuranov \& Postnov 2006), 

\begin{eqnarray}
D(R) = \left\{ \begin {array}{ll}
1,                   & R < R_{\rm c} \\
(R_{\rm c}/R)^2,     & R_{\rm c} < R < R_{\rm h} \\
(R_{\rm c}R_{\rm h})^2/R^4,              & R_{\rm h} < R < R_{\rm t}, 
\end{array} \right.
\label{eq2a}
\end{eqnarray}

\noindent
we calculate the energy density of stellar photons inside the cluster core from the formula,
\begin{eqnarray}
U_{\rm rad} = {{L_{\rm GC}}\over{c R_{\rm t}^2}} 
{{18R_{\rm t}^2 - 3\sqrt{6R_{\rm t}^3R_{\rm c}} -2R_{\rm c}^2}
\over{6(\sqrt{6R_{\rm t}R_{\rm c}^3} - 2R_{\rm c}^2)}},
\label{eq2b}
\end{eqnarray}
\noindent
where $R_{\rm c}$ is the GC core radius; $R_{\rm t}$ is its tidal radius, and the half mass radius, $R_{\rm h}$, is related to those both by $R_{\rm h} = \sqrt{2R_{\rm c}R_{\rm t}/3}$.
For typical parameters, $R_{\rm c}=0.5$ pc, $R_{\rm t}=50$ pc, and $L_{\rm GC}=10^5$ L$_\odot
 = 3\times 10^{38}$ erg s$^{-1}$, we obtain $U_{\rm rad}\approx 300$ eV cm$^{-3}$.
The energy density of this stellar radiation field clearly dominates over the energy density of the microwave background radiation (MBR), $U_{\rm MBR} = 0.25$ eV cm$^{-3}$. It can also dominate over the 
energy density of the magnetic field inside the GC provided that the magnetic field is not
stronger than $B_{\rm GC}\sim 10^{-4}$ G. This limit has been obtained assuming the equipartition between the energy density of stellar photons and the energy density of the magnetic field. Since such strong magnetic field inside the GCs seems unlikely
(see estimate given by Eq.~\ref{eq2} for the region of the pulsar wind shock), we conclude that leptons injected by the pulsars in the GC cores lose energy mainly on IC scattering of the stellar and microwave radiation.

\section{Acceleration of leptons inside globular clusters} 

Let us estimate the maximum energies which leptons ($e^\pm$) can reach in such pulsar wind shocks. The obvious limit is introduced by the condition that the Larmor radius of accelerated particles can not be greater than  the dimension of the shocks, i.e. $R_{\rm L} = cp/(eB)<R_{\rm sh}$, where $c$ is the velocity of light, $e$ is the charge of an electron, $p$ is the momentum of the particle, and $B$ is the magnetic field strength at the shock. For the above estimated magnetic field at the shock, we get  
\begin{eqnarray}
E_{\rm sh}^{\rm max} < 1.4\times 10^3\sqrt{\sigma}B_9/P_4^2~~~{\rm TeV}.
\label{eq3}
\end{eqnarray}
Additional limits can be constrained by the energy losses of leptons on the synchrotron and inverse Compton (IC) processes during their acceleration. In order to estimate these limits we compare the acceleration rates of leptons with their energy loss rates on the mentioned processes.
The acceleration rate is independent on the energy of lepton $E$ provided that so called acceleration coefficient $\xi$ is constant. It can be written as, 
\begin{eqnarray}
\left({{dE}\over{dt}}\right)_{\rm acc} = \xi  c E/R_{\rm L}\approx 
10^{13}\xi B_{\rm sh}~~~{\rm eV~s^{-1}},
\label{eq4}
\end{eqnarray}
\noindent
and the corresponding acceleration time scale,
\begin{eqnarray}
t_{\rm acc} = E/(dE/dt)_{\rm acc}\approx 10^5 E_{\rm TeV}/(\xi B_{-6})~~~{\rm s}.
\label{eq4a}
\end{eqnarray}
\noindent
where $E = 1 E_{\rm TeV}$ TeV is the energy of leptons, $B_{\rm sh} = 10^{-6}B_{-6}$ G is the magnetic field strength at the shock, and $c$ is the velocity of light.

On the other hand, the synchrotron energy loss rate is,
\begin{eqnarray}
\left({{dE}\over{dt}}\right)_{\rm syn} = {{4}\over{3}}c\sigma_{\rm T}U_{\rm B}\gamma_{\rm e}^2\approx 
8.3\times 10^{-4}B_{\rm sh}^2\gamma_{\rm e}^2~~~{\rm eV~s^{-1}},
\label{eq5}
\end{eqnarray}
\noindent
where $U_{\rm B}$ is the energy density of magnetic field, $\sigma_{\rm T}$ is the Thomson cross section, and $\gamma_{\rm e}$ is the Lorentz factor of leptons. 
By comparing Eqs.~\ref{eq4} with~\ref{eq5} and applying Eq.~\ref{eq2} we get the limit on the maximum energy of accelerated leptons due to the synchrotron losses,
\begin{eqnarray}
E_{\rm syn}^{\rm max}\approx 4\times 10^2P_4 \xi_{-2}^{1/2}/(\sigma^{1/4}B_9^{1/2})~~~{\rm TeV.} 
\label{eq6}
\end{eqnarray}
\noindent
where $\xi = 0.01\xi_{-2}$. 
This limit is more restrictive than the limit due to the dimension of the shock ($E^{\rm max}_{\rm syn} < E^{\rm max}_{\rm sh}$) provided that the following condition is fulfilled, 
\begin{eqnarray}
P_4 <  1.5\sigma^{1/4}B_9^{1/2}/\xi_{-2}^{1/6}.
\label{eq7}
\end{eqnarray}

Leptons accelerated at the shock are also immersed in the strong radiation field coming from the stellar population within the GC. Above we estimated the average density of stellar photons in the core of GC on $U_{\rm rad}\approx 300$ eV cm$^{-3}$. 
This allows us to estimate the energy loss rate on ICS in the Thomson (T) regime,
\begin{eqnarray}
\left({{dE}\over{dt}}\right)_{\rm IC}^{\rm T} = {{4}\over{3}} c\sigma_{\rm T}U_{\rm rad}\gamma_{\rm e}^2\approx  7.6\times 10^{-12}\gamma_{\rm e}^2~~~{\rm eV~s^{-1}}.
\label{eq8}
\end{eqnarray}
The Thomson regime is valid only for leptons with energies below $E_{T/KN}\approx 100$ GeV.
By comparing Eqs.~\ref{eq4} with~\ref{eq8} and applying the limit $E_{\rm T/KN}$, we obtain
that the saturation of acceleration of leptons can occur due to IC losses in the Thomson regime provided that $\xi < \xi_{\rm crit} = 10^{-7}/B_{-6}$. However, this case is not specially interesting from the point of view of TeV $\gamma$-ray production. Note that the acceleration of particles at relativistic shocks (defined by $\xi$) is
usually considered as much more efficient. For example, observations of the TeV $\gamma$-ray emission from the Crab Nebula show that the acceleration of particles occurs during a few Larmor cycles. Thus, $\xi$ has to be of the order of $\sim 0.1-1$.
Therefore, here we consider the case of saturation of lepton acceleration in the Klein-Nishina (KN) regime.
The energy loss rate in the KN regime depends only logarithmically on energy of leptons, i.e. depends very weakly on energy of leptons.
So then, if the acceleration coefficient is significantly larger than $\xi_{\rm crit}$ (as considered in this paper for the pulsar wind relativistic shocks) the saturation of acceleration of leptons might be caused either by synchrotron energy losses or dimension of the shock (depending on the parameters of the specific pulsar, see Eq.~\ref{eq7}). 

However, the limit on the maximum energies of accelerated leptons can be also given by the advection time scale of leptons along the surface of the shock. The velocity of the plasma flow downstream of the relativistic shock can be estimated at $v_{\rm adv}\sim c/3$. Then, the advection time scale is, 
\begin{eqnarray}
t_{\rm adv} = R_{\rm sh}/v_{\rm adv} = 0.5R_{\rm c}/(N_{\rm p}^{1/3}v_{\rm adv}).
\label{eq8a}
\end{eqnarray}
\noindent
where $R_{\rm sh}$ is defined by Eq.~\ref{eq1b}. By comparing the acceleration time scale of leptons (Eq.~\ref{eq4a}) with their advection time scale along the shock (Eq.~\ref{eq8a}),
we obtain another limit on the maximum energies of leptons,
\begin{eqnarray}
E_{\rm adv}^{\rm max} = 1.5\xi_{-2} B_{-6}R_{\rm pc}/(N_{100}^{1/3}v_{\rm c})~~~{\rm TeV,}
\label{eq8b}
\end{eqnarray}
\noindent
where $R_{\rm c} = 1R_{\rm pc}$ pc, $N_{\rm p} = 100N_{100}$, and $v_{\rm c} = v_{\rm adv}/c$. For likely parameters of the considered scenario, $B_{-6} = 1$, $R_{\rm pc} = 0.5$, $N_{100} = 0.3$, and $v_{\rm c} = 0.3$, we estimate the maximum energies of accelerated leptons due to their advection along the shock at
$E_{\rm adv}^{\rm max}\approx  370\xi$ TeV. For reasonable acceleration efficiency at relativistic shock of the order of $\xi\sim 0.1-0.01$, this maximum energies of leptons are limited to the range $E_{\rm adv}^{\rm max}\sim 4-40$ TeV. Therefore, we conclude that the acceleration process of leptons is limited by their advection process along the surface of the shock.

Based on the above general considerations, we conclude that  the maximum energies of leptons accelerated at the shocks are likely to be limited by advection process of leptons along the shock and not by the
dimension of the shock in the pulsar wind $E_{\rm sh}^{\rm max}$ or the energy losses
on the synchrotron or IC scattering processes. Leptons can be accelerated inside  GC to energies which allow them to comptonize the stellar and MBR photons to the TeV $\gamma$-ray energy range.
 
Leptons injected with different energies in the central core of the globular cluster diffuse gradually in the outward direction. We consider the central region with the radius $R_{\rm h}$ in which half of the mass of the GC is contained.  Assuming the Bohm diffusion, the  average time spend by leptons with energy E inside this region is,   
\begin{eqnarray}
t_{\rm diff} = R_{\rm h}^2/D_{\rm diff},
\label{eq9}
\end{eqnarray}
\noindent
where $D_{\rm diff} = R_{\rm L}c/3$ is the diffusion coefficient, $R_{\rm L} =ep/eB_{\rm GC}\approx 3\times 10^{15} E_{\rm TeV}/B_{-6}$  cm is the Larmor radius of leptons in the magnetic field of the GC, and $B_{\rm GC} = 10^{-6}B_{-6}$ G is the average magnetic field in the GC.  

In order to answer the question whether leptons can lose significant part of their energy during the diffusion through
the thermal radiation field inside the GC we estimate the collision rate on IC process in the Thomson regime for supposed parameters inside the GC. Let us use the following parameters, $E_{\rm TeV} = 0.1$, $B_{\rm GC} = 3\times 10^{-6}$ G, $R_{\rm h} = 3$ pc, and the average density of thermal photons inside the GC, $N_{\rm ph} = 100$ cm$^{-3}$. Then, $R_{\rm L} = 10^{14}$ cm, $D_{\rm diff} = 10^{24}$ cm$^2$ s$^{-1}$, and $t_{\rm diff} = 8\times 10^{13}$ s. For this diffusion time scale, the collision rate of leptons is of the order of $N_{\rm col} = c t_{\rm diff} \sigma_{\rm T} N_{\rm ph}\approx 160$, i.e. leptons might interact up to a hundred times before leaving the globular cluster. Note that the diffusion time and the ICS cross section in the KN regime are inversely proportional to energy of leptons. Therefore, leptons with energies up to $\sim 1$ TeV should interact frequently with stellar photons producing $\gamma$-rays with comparable energies. However, leptons with energies above $\sim 1$ TeV have chance to escape from the radiation field of the GC without interaction with stellar photons. On the other hand, these high energy leptons interact frequently with the MBR producing $\gamma$-rays in the energy range from tens MeV up to hundreds GeV. In the next section, we perform more detailed calculations of the $\gamma$-ray spectra produced in the process considered above by assuming injection of leptons with the mono-energetic and the power-law spectra with different spectral indexes, injection efficiencies, and high energy cut-offs.

\section{Gamma-ray production inside globular clusters} 

\begin{figure}
\vskip 5.2truecm
\includegraphics{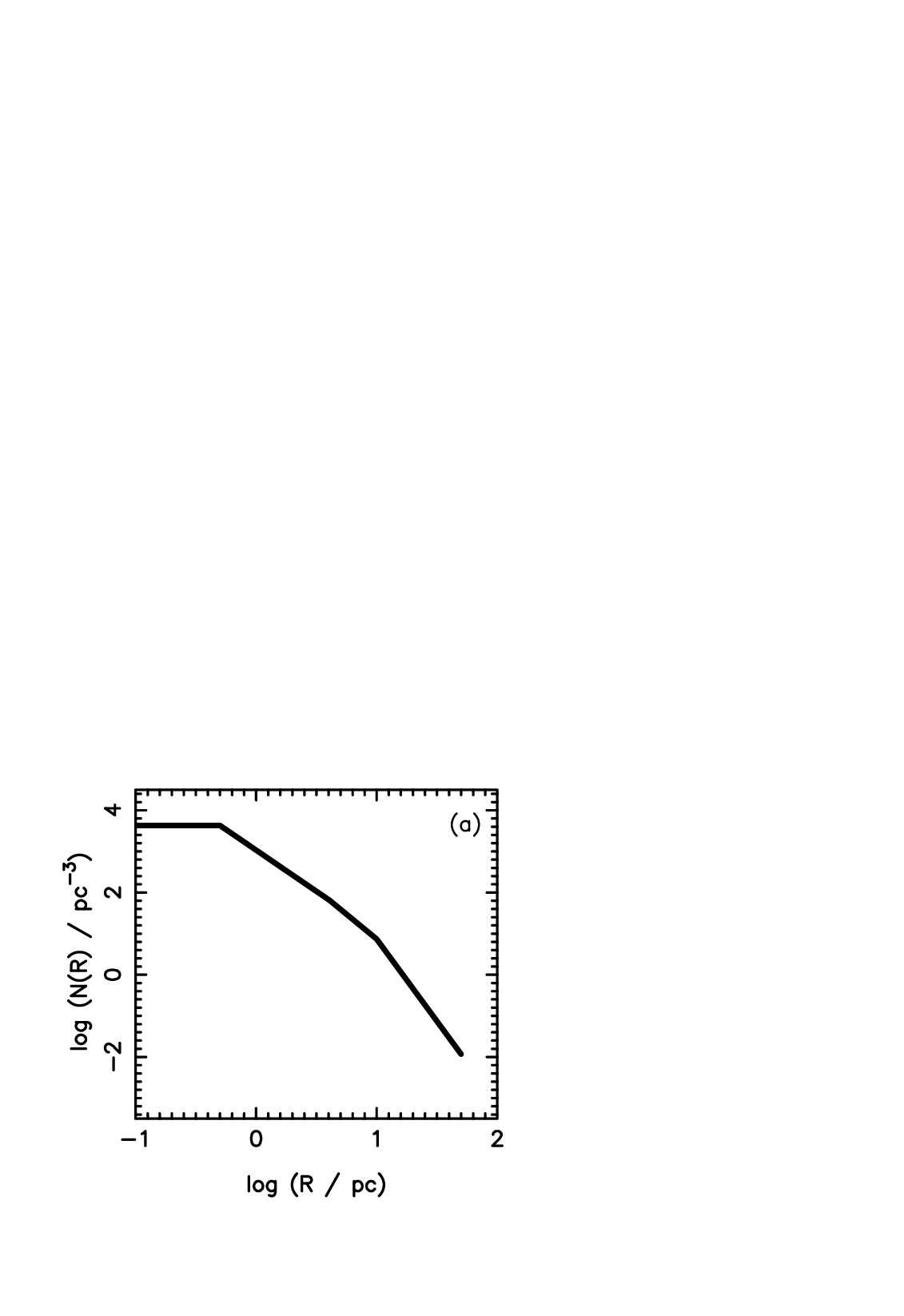}
\includegraphics{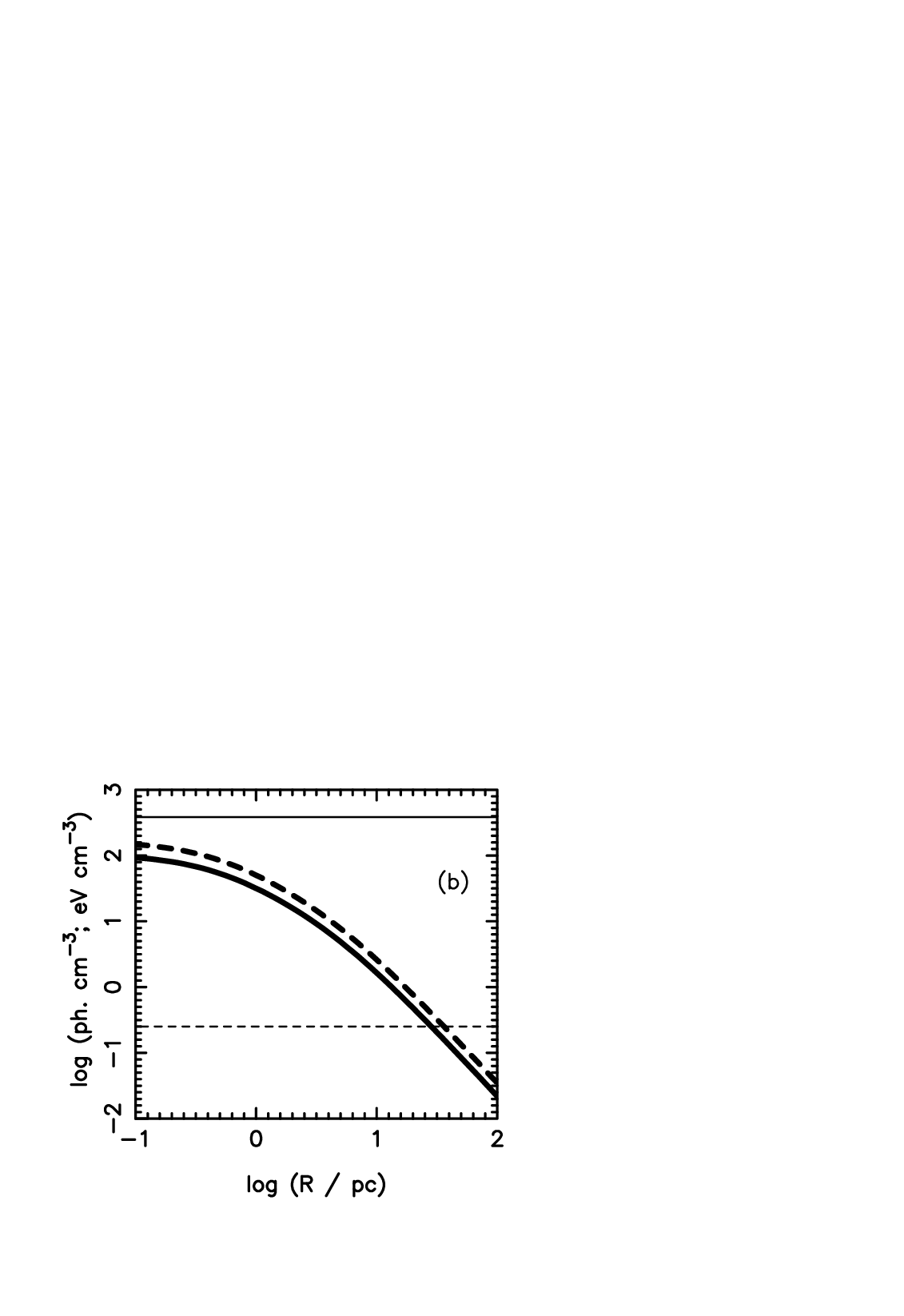}
\caption{The density of stars (figure a), and density and energy density of stellar photons (b) as a function of distance $R$ from the center of a typical globular cluster with the mass $10^5$ M$_\odot$, the core radius $R_{\rm c} = 0.5$ pc,  the half mass radius $R_{\rm h} = 4$ pc, the tidal radius $R_{\rm t} = 50$ pc. The stellar density profile is defined by Eq.~\ref{eq2a}. The photon densities and energy densities of stellar photons are shown by the thick solid and dashed curves and the corresponding values for the microwave background radiation are shown by the thin horizontal lines.}
\label{fig1}
\end{figure}

Let us assume that the part, $\eta$, of the wind energy of pulsars inside the GC is converted into relativistic leptons (so called energy conversion efficiency). Then
using Eq.~\ref{eq1}, the power injected in relativistic leptons is,
\begin{eqnarray}
L_{\rm e} = \eta N_{\rm p} L_{\rm p} = 
1.2\times 10^{35}\eta N_{\rm p}B_9^2P_4^{-4}~~~{\rm erg~s^{-1}}.
\label{eq9}
\end{eqnarray}
\noindent
The spectrum of leptons accelerated at the shocks can be approximated by a single power-law with the spectral index $\alpha$ between $E_{\rm min}$ and $E_{\rm max}$. 
These leptons produce $\gamma$-rays as a result of ICS of thermal and MB photons. In order to perform numerical calculations of the $\gamma$-ray spectra, first we have to define the radiation field inside GC as a function of the distance from its center. The density profile for the distribution of stars inside GC (Eq.~\ref{eq2a}) is normalized to the total number of stars inside GC which is typically in the range $N_\star^{\rm tot} = 10^5-10^6$ M$_\odot$. The density of stars as a function of the distance is 
$N_\star(R) = A D(R) dR$, where $A = N_\star^{\rm tot}/[2R_{\rm c}-2R_{\rm c}^2/(3R_{\rm h})-
R_{\rm c}^2R_{\rm h}^2/(3R_{\rm t}^3)]\approx N_\star^{\rm tot}/(2R_{\rm c})$ and $D(R)$ is given by Eq.~\ref{eq2a}. As a first approximation, it is assumed that the average star inside GC has the basic parameters close to that of our Sun, i.e. the surface temperature $6000$ K and the radius $7\times 10^{10}$ cm (note that the mass to light ratio for typical GC is close to $\sim$2). 
The density of stellar photons at a specific distance from the center of GC is calculated by integration over the whole distribution of stars inside the GC. The examples of the density profiles for distribution of stars and the density and energy density of stellar photons inside the GC with the luminosity  $L_{\rm GC} = 10^5$ M$_\odot$ are shown in Fig.~\ref{fig1}. 
Inside the globular cluster core, defined by $R_{\rm c} = 0.4$ pc, the density of photons drops slowly with the distance from the core reaching the value of approximately a factor of two lower at $R_{\rm c}$ than at the center of the core. Outside the half mass radius, defined by $R_{\rm h} = 4$ pc, density of photons drops almost proportionally to $\propto R^{-2}$. For comparison we also show the photon and photon energy densities of MBR. Note, that in the case of considered GC, the photon density of the MBR dominates over density of stellar photons. In contrast, the photon energy density of MBR is much lower than the stellar photon energy density in the whole volume of the GC. Therefore low energy leptons lose mainly energy on scattering stellar radiation in the Thomson regime. However leptons with large energies lose energy mainly by scattering the MBR in the Thomson regime since their scattering of stellar radiation is strongly suppressed due to the KN effect.

\subsection{Injection of mono-energetic leptons}

We calculate the $\gamma$-ray spectra produced by leptons which IC up-scatter the stellar photons and the MBR. For considered energies of leptons the scattering process of stellar radiation can occur in the T or the KN regimes.
We neglect the synchrotron energy losses of leptons for the considered range of their energies and values of the magnetic field inside the GCs. 
It is assumed that leptons are injected homogeneously inside the core of GC. 
Since diffusion of leptons from the core occurs in the variable thermal radiation field, the importance of the scattering process in both considered radiation
fields (thermal photons and MBR) can differ significantly. In order to calculate the $\gamma$-ray spectra produced by leptons diffusing outward the center of GC we apply the Monte Carlo method.
The optical depth for leptons on the ICS is determined during their gradual diffusion in variable radiation field. Therefore, the distance from the core of GC at which lepton 
with specific energy $E$ produces $\gamma$-ray photon is simulated numerically. For known location of the interaction place inside the GC, we simulate the energy of $\gamma$-ray photon produced in the IC process. The energy of primary lepton is reduced by the energy of produced $\gamma$-ray and the process of simulation is repeated for the next interaction place and energy of produced $\gamma$-ray photon. The propagation process of leptons is considered up to the distance of 100 pc from the center of the GC.
At this distance, the contribution of stellar photons becomes unimportant. All secondary $\gamma$-rays produced by leptons are sorted in specific energy ranges in order to obtain the spectrum of $\gamma$-rays produced inside the GC. The method described above allows us to calculate the spectra of produced $\gamma$-rays as a function of distance from the center of GC. Therefore, we can estimate the extent of the specific $\gamma$-ray source on the sky related to the GC.

\begin{figure}
\vskip 9.5truecm
\includegraphics{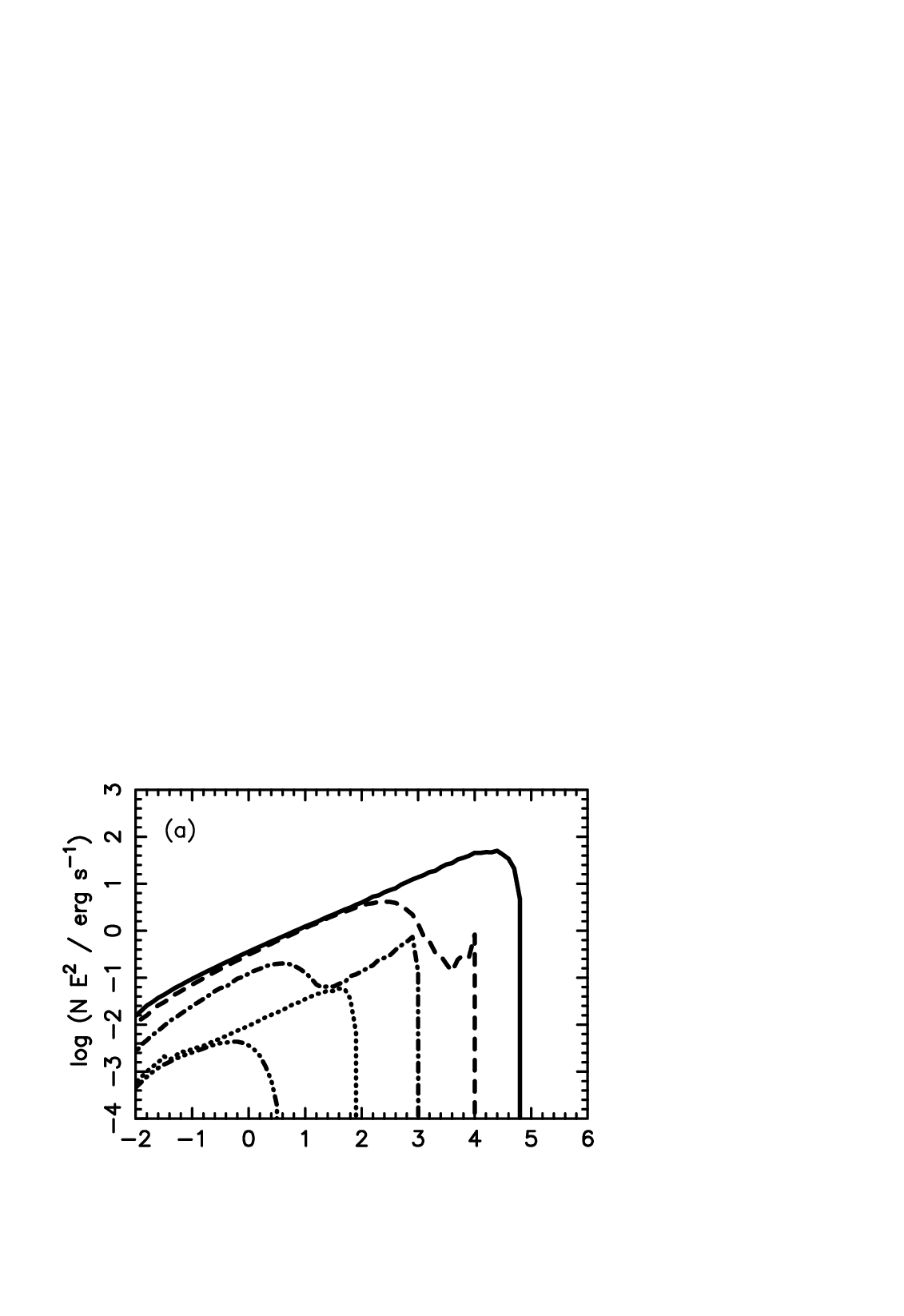}
\includegraphics{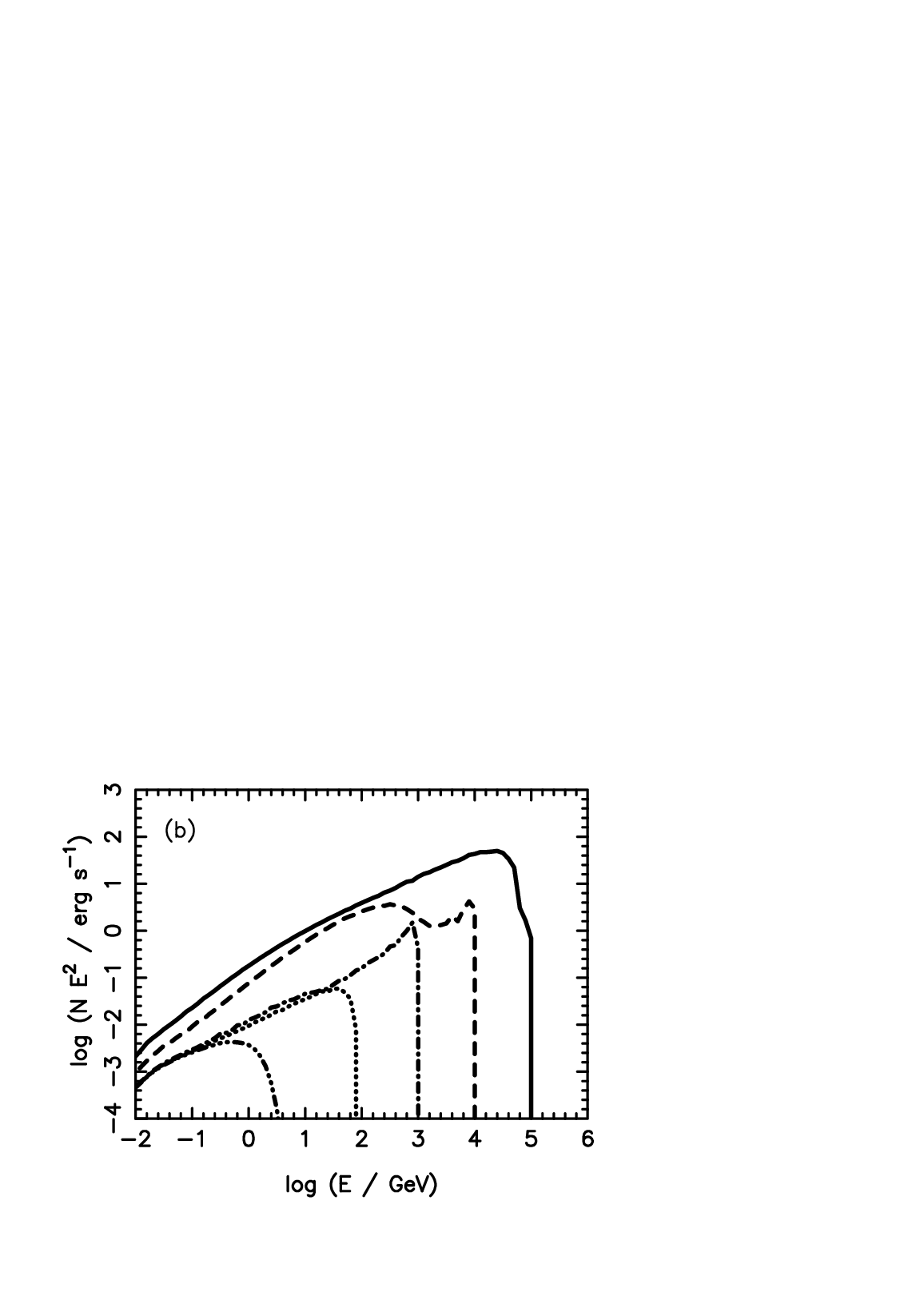}
\caption{Differential $\gamma$-ray spectra (multiplied by the energy squared) produced in the IC scattering of the stellar and MBR radiation by mono-energetic leptons with energies $E = 10$ GeV (triple-dot-dashed curve), $10^2$ GeV (dotted), $10^3$ GeV (dot-dashed), $10^4$ GeV (dashed), and $10^5$ GeV (solid), after normalization to a single lepton per second. Leptons are injected by the millisecond pulsars in the core of GC and diffuse in the outward direction in the magnetic and radiation field created by the stars and the MBR. The magnetic field strength is assumed equal to $B_{\rm GC} = 10^{-6}$ G.  The diffusion process of leptons is followed up to 100 pc from the core of GC. The total stellar luminosity of GC is equal to $L_{\rm GC} = 10^5L_\odot$ (a) and $10^6L_\odot$ (b).}
\label{fig2}
\end{figure}

In order to perform specific calculations we assume that leptons are injected 
homogeneously in the core of GC. Differential $\gamma$-ray spectra (multiplied by the energy squared), calculated for mono-energetic leptons injected at the core of GC, are shown in Fig.~\ref{fig2} for the total luminosity of the GC $L_{\rm GC} = 10^5L_\odot$ (figure a) and $10^6L_\odot$ (b) after normalization to a single injected lepton per second.
For leptons with low energies ($\le 10^2$ GeV), the IC scattering of stellar photons and MBR occurs only in the T regime. As we noted above the photon density of MBR dominates over the density of stellar photons for $L_{\rm GC} = 10^5L_\odot$, but the energy density of the MBR is significantly lower. Therefore, such leptons lose energy mainly by scattering stellar photons. For the total luminosity of GC equal to $L_{\rm GC} = 10^6L_\odot$ both the photon density and energy density of stellar photons dominate over the MBR photon and energy density. However, when the energies of injected leptons are larger, the situation becomes more complicated due to the influence of the ICS in the KN regime. 
Then, for sufficiently large energies of leptons, the scattering process of MBR can dominate the energy losses of leptons.

Let us at first consider the case of the GC with luminosity $L_{\rm GC} = 10^5L_\odot$. 
It is clear from our calculations that scattering of MBR can dominate energy loss process of leptons. 
Already for leptons with energies $10^3$ GeV, the power which goes into the $\gamma$-rays due to scattering of MBR in the T regime and the stellar radiation in the KN regime becomes comparable (see the $\gamma$-ray spectrum marked by the dot-dashed curve in Fig.~\ref{fig2}a with the characteristic lower energy 'bump' due to IC scattering of MBR in the T regime and the characteristic 'peak' at the highest energies due to the IC scattering of the stellar radiation in the KN regime).
For the largest considered energies of leptons ($10^5$ GeV) only the $\gamma$-ray spectrum produced in the IC scattering process of the MBR is clearly seen but the scattering of stellar photons is completely suppressed due to the KN regime.
However, when the density of stellar photons dominates over the density of the MBR (the case of the GC with the total luminosity equal to $L_{\rm GC} = 10^6L_\odot$), then contribution to the $\gamma$-ray spectrum from leptons which scatter also stellar photons in the KN regime becomes evident up to the largest considered energies of leptons (see characteristic bumps at the end of $\gamma$-ray spectrum due to scattering in the KN regime, Fig.~\ref{fig2}b). Note, that in case of such luminous GCs, the $\gamma$-ray spectra produced by leptons with energies above $10^4$ GeV are clearly dominated by photons produced in the scattering process of the MBR. For leptons with energies $10^3$ GeV, the contribution from scattering of the stellar and the MBR photons becomes comparable but for leptons with energies below $10^2$ GeV, $\gamma$-ray spectrum is formed mainly by scattering the stellar radiation since density of these photons in the core of the GC dominates over the density of the MBR by a factor of $\sim$3.

\subsection{Injection of leptons with the power law-spectra}

As a next step, we analyze the features of the $\gamma$-ray spectra produced by leptons with the power-law spectrum. As we discussed above (Sect. ~3), leptons are accelerated with the power-law spectrum at the relativistic shocks which are created in collisions of the pulsar winds inside the core of GCs. The maximum energies of leptons, $E_{\rm max}$, are determined by the radiation and escape processes of leptons and by the efficiency of their acceleration. For illustration purposes, we consider in this section the cases with $E_{\rm max}$ = 3, and 30 TeV which are consistent with the above derivations (Sect.~3). Since  details of the acceleration process are not precisely known, we consider a range of the injection spectra with the spectral indexes between $\alpha$=2-3. On the other hand, we also fix the low energy cut-off in these power-law spectra on $E_{\rm min} = 1$ GeV or 100 GeV. This low energy cut-off might correspond to energies of leptons injected into the pulsar wind from the inner magnetopsheres of the millisecond pulsars. The models of the radiation processes in the inner magnetospheres of MSPs show that if the acceleration of leptons occurs as expected in terms of a space charge limited flow model (developed by Muslinov \& Tsygan~1992), then their acceleration can be limited by the curvature energy losses to energies $\sim$1-10 TeV (e.g. Bulik, Rudak \& Dysk~2000; Luo, Shibata, Melrose~2000; Harding, Usov \& Muslimov~2005). In fact, this possibility is investigated in this paper in the case of injection of mono-energetic leptons (see Sect~4.1 and 5.5). However, the processes occurring in the inner magnetospheres of MSPs are not well known. 
If the temperature of the polar cap region is sufficiently large, above $\sim 3\times 10^6$ K, then plenty of $e^\pm$ pairs can appear in the acceleration region due to $\gamma$-ray production in ICS process and $\gamma$-ray photon absorption in the magnetic field or in two photon collisions (Harding \& Muslinov~2002; Harding, Muslinov \& Zhang~2002). These $e^\pm$ pairs may saturate the electric field in the acceleration region. As a result, leptons may escape from the inner magnetosphere with relatively low energies.

The power-law spectrum of leptons is normalized to the part of power of the millisecond pulsars winds present inside the core of GC, $L_e$ (see Eq.~\ref{eq9}), assuming the average pulsar parameters, $B_9 = 1$ and $P_4 = 1$. The energy conversion efficiency of the pulsar energy loss rate to relativistic leptons is taken equal to $\eta = 1\%$, and the number of millisecond pulsars in the GC core is equal to  $N_{\rm p} = 30$.
In order to perform the Monte Carlo simulations, we divide the spectrum of leptons into the narrow energy bins ($E, E+\Delta E$) since the diffusion and ICS processes of leptons  strongly depends on their energy.

\begin{figure}
\vskip 9.5truecm
\includegraphics{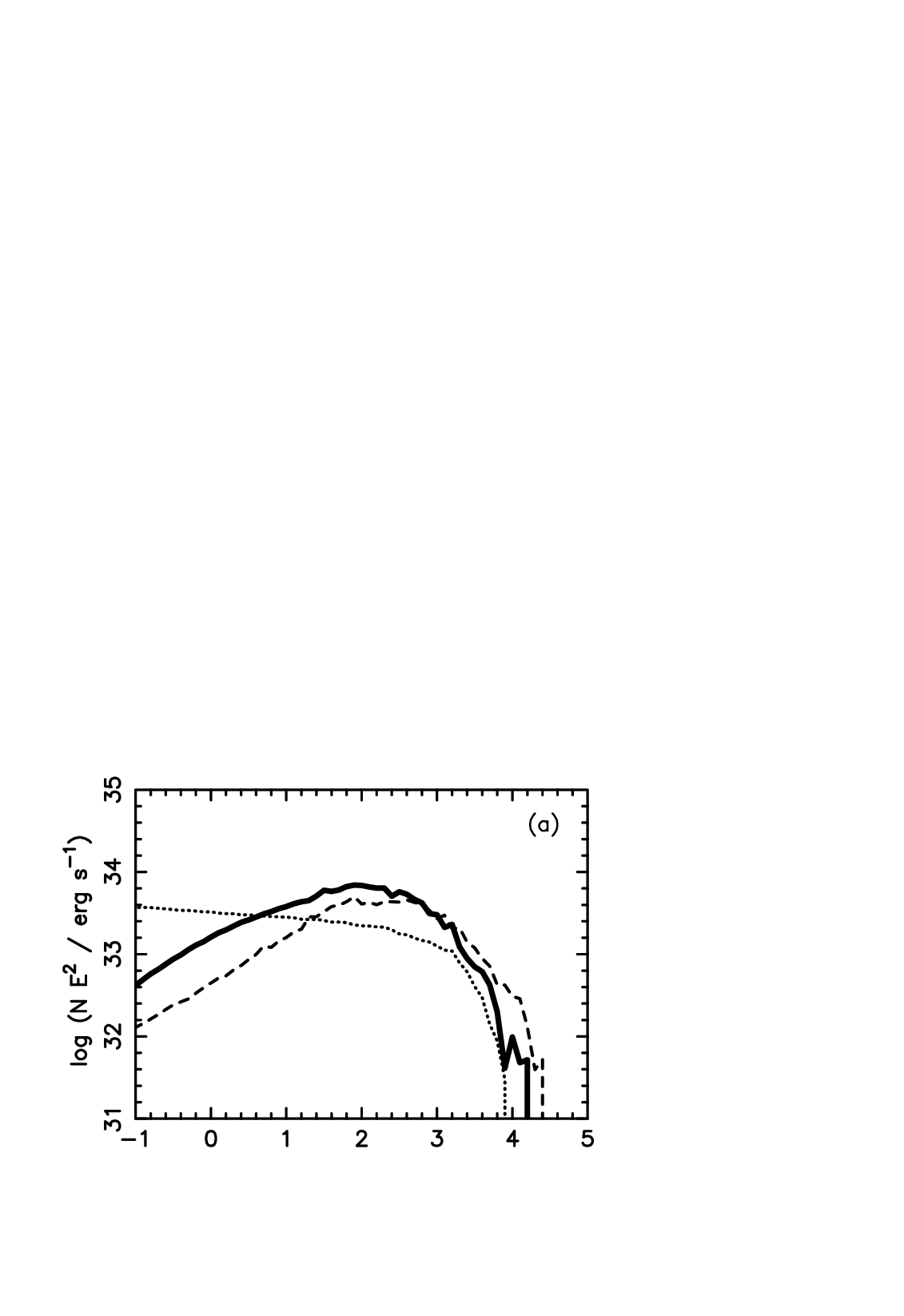}
\includegraphics{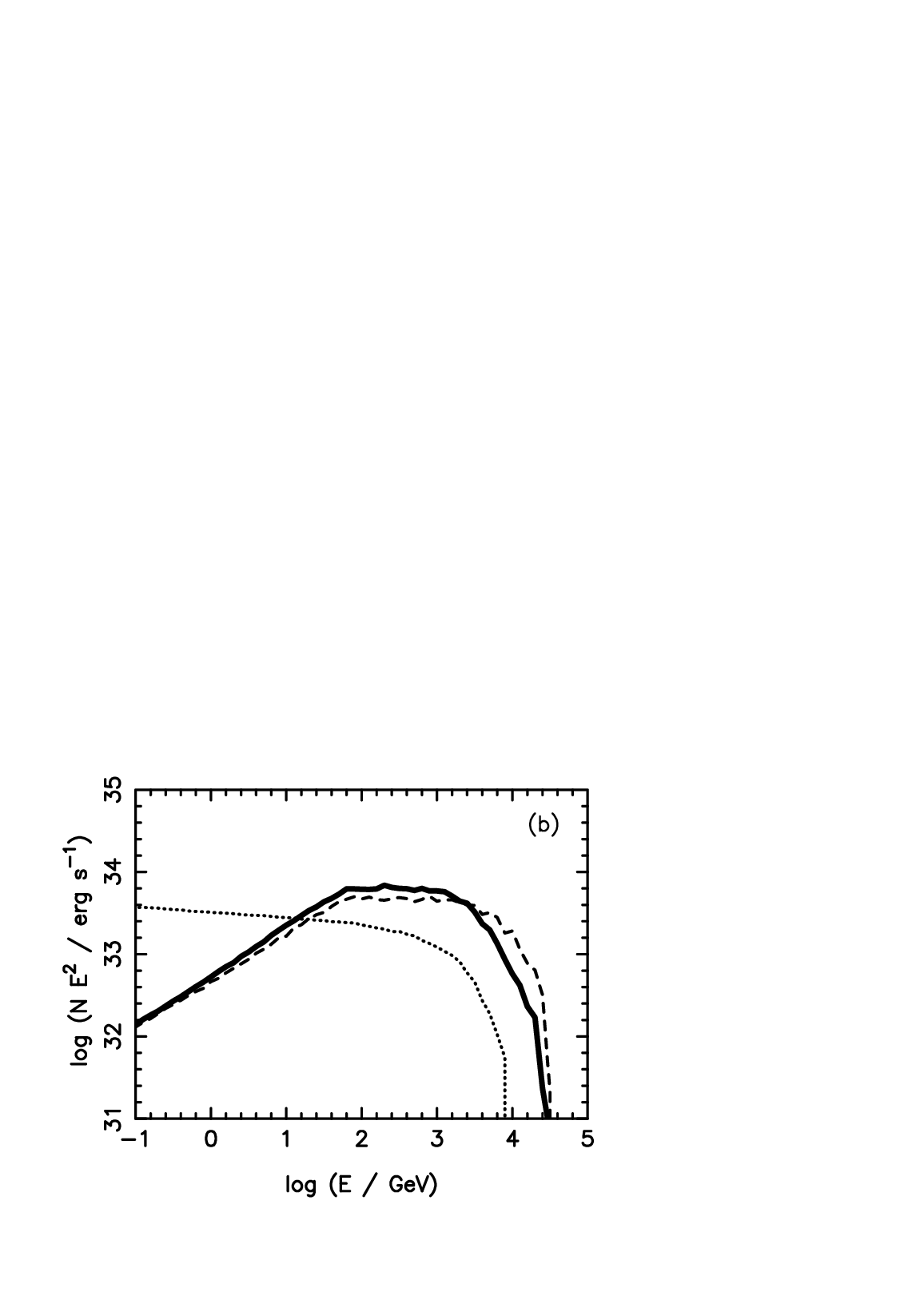}
\caption{Differential $\gamma$-ray spectra (multiplied by the energy squared) 
produced by leptons with the power law spectrum and spectral index $\alpha = 2.1$ (extending between $E_{\rm min} = 10^2$ GeV and $E_{\rm max} = 3\times 10^4$ GeV) in the presence of only: MBR (dotted curve), stellar radiation (dashed), and both MBR and stellar radiations (solid). The spectrum of leptons is normalized to the power of 30 millisecond pulsars
(see Eq.~\ref{eq9}), the energy conversion efficiency is assumed equal to $\eta = 0.01$, and the average pulsar parameters are $B_9 = 1$ and $P_4 = 1$. The results for the the total stellar luminosity of the GC equal to $L_{\rm GC} = 10^5L_\odot$ are shown in figure (a) and for $10^6L_\odot$ in figure (b). The magnetic field strength inside the GC is fixed on $B_{\rm GC} = 10^{-6}$ G.}
\label{fig3}
\end{figure}
\begin{figure}
\vskip 9.5truecm
\includegraphics{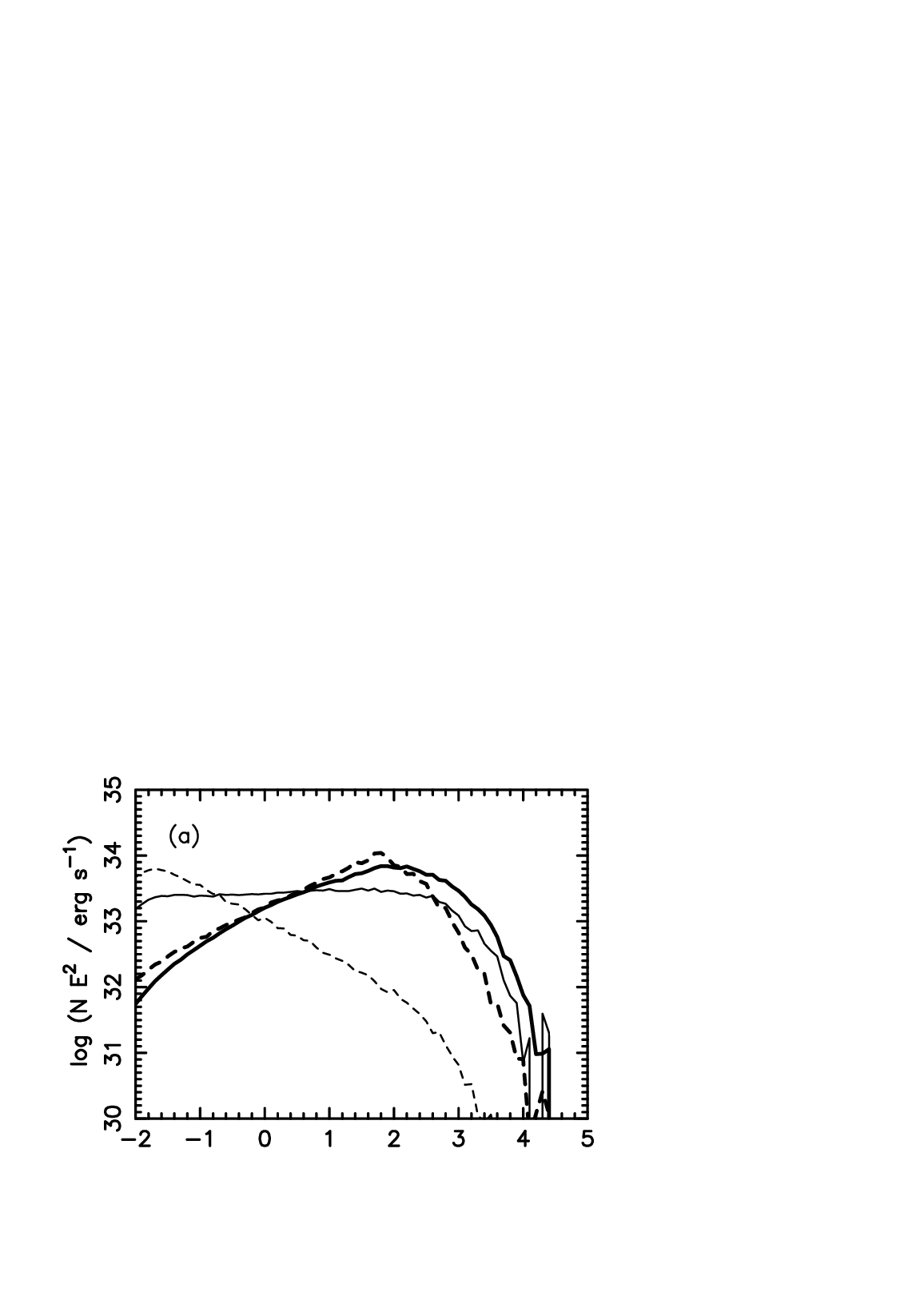}
\includegraphics{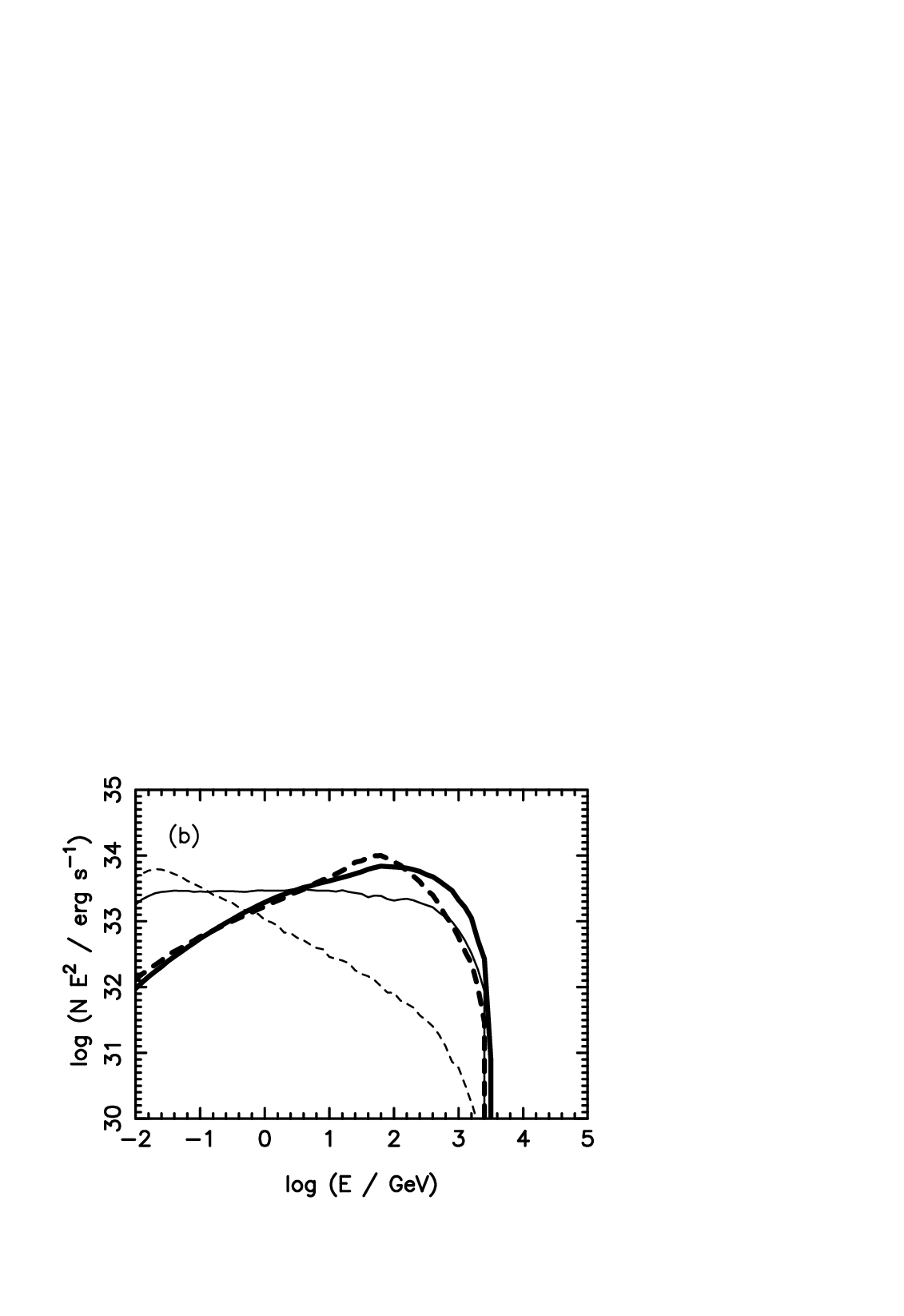}
\caption{The differential spectra of $\gamma$-rays (multiplied by the energy squared) produced by
leptons with the power law spectrum and spectral indices $\alpha = 2.1$ (solid curves), and 
$3$ (dashed). The spectrum of leptons extends between $E_{\rm min} = 100$ GeV (thick curves)
or $1$ GeV (thin curves) and $E_{\rm max} = 30$ TeV (a) or  $3$ TeV (b). The spectra of leptons are normalized as in Fig.~\ref{fig3}.
The GC with the luminosity $L_{\rm GC} = 10^5$ M$_\odot$ is considered.
}
\label{fig4}
\end{figure}

To understand better the $\gamma$-ray production process, we calculate the differential  $\gamma$-ray spectra produced by leptons in the IC scattering process of the MBR only and stellar radiation only (see dotted and dashed curves in Figs.~\ref{fig3}a and b). We also show the $\gamma$-ray spectra produced in the presence of both radiation fields (solid curves in Figs.~\ref{fig3}). 
The power-law injection spectrum of leptons with spectral index equal to $\alpha = 2.1$, extending between $E_{\rm min} = 100$ GeV and $E_{\rm max} = 3\times 10^4$ GeV, is investigated. We follow diffusion of leptons up to the distance of 100 pc from the core of GC. Note that since the scattering of MBR by leptons occurs only in the T regime, the   
$\gamma$-ray spectrum, obtained in the case of the MBR only, has the shape characteristic for completely cooled leptons and the power-law $\gamma$-ray spectrum with spectral index close to $(\alpha + 2)/2 = 2.05$. The $\gamma$-ray spectra obtained from IC scattering of stellar radiation only are completely different. As already discussed in Sect.~4.1, leptons with energies above $\sim$0.1 TeV interact with stellar photons in the KN regime. Some of them can escape from the region with radius 100 pc around the core of GC without interaction. Therefore, the highest energy part of the $\gamma$-ray spectrum becomes steeper and this effect is stronger for less luminous GCs. 
The $\gamma$-ray spectra obtained in the case of both photon fields (stellar and MBR) are combination of the above described spectra.
In the case of the GCs with low luminosity ($L_{\rm GC} = 10^5L_\odot$, Fig.~\ref{fig3}a), leptons scattering the MBR contribute mainly to the lower energy part of the $\gamma$-ray spectrum (see differences between dashed and full curves in Fig.~\ref{fig3}a). If the luminosity of GC is large ($L_{\rm GC} = 10^6L_\odot$, Fig.~\ref{fig3}b), the contribution to the $\gamma$-ray spectrum at low energies from scattering of the MBR is negligible since the energy and photon density of stellar radiation completely dominates over the energy and stellar density of the MBR. The scattering of the MBR by leptons with the highest energies has some small effect on the produced spectrum of $\gamma$-rays (shift
to lower energies in respect to the $\gamma$-ray spectra from scattering of stellar radiation only due to the additional energy losses on the MBR).  

We also consider the dependence of the $\gamma$-ray spectra on the parameters describing the injection spectrum of leptons (Figs.~\ref{fig4}a and b).
For the GC with luminosity $L_{\rm GC} = 10^5$ M$_\odot$,
we calculate the $\gamma$-ray spectra for two spectral indices of leptons ($\alpha = 2.1$ and 3), two low energy cut-offs ($E_{\rm min} = 1$ and 100 GeV), and two high energy cut-offs ($E_{\rm max} = 3$ and 30 TeV). As in the previous cases, the diffusion of leptons up to 100 pc from the center of GC is considered. 
The $\gamma$-ray spectra peak at the GeV or the TeV energy ranges depending on $E_{\rm min}$. Therefore, the broad band $\gamma$-ray observations, through GeV-TeV energy range, with the GLAST and Cherenkov telescopes (MAGIC, HESS, VERITAS) should confirm (or disprove) models considered here for different spectra of injected leptons. 
The detection (or non-detection) of $\gamma$-ray emission from GCs should put a limit on the parameters of the acceleration models of leptons in inner magnetopsheres of millisecond pulsars and/or at the pulsar wind relativistic shocks.

\subsection{Dependence on the distance from the center of Globular Cluster}

\begin{figure}
\vskip 9.5truecm
\includegraphics{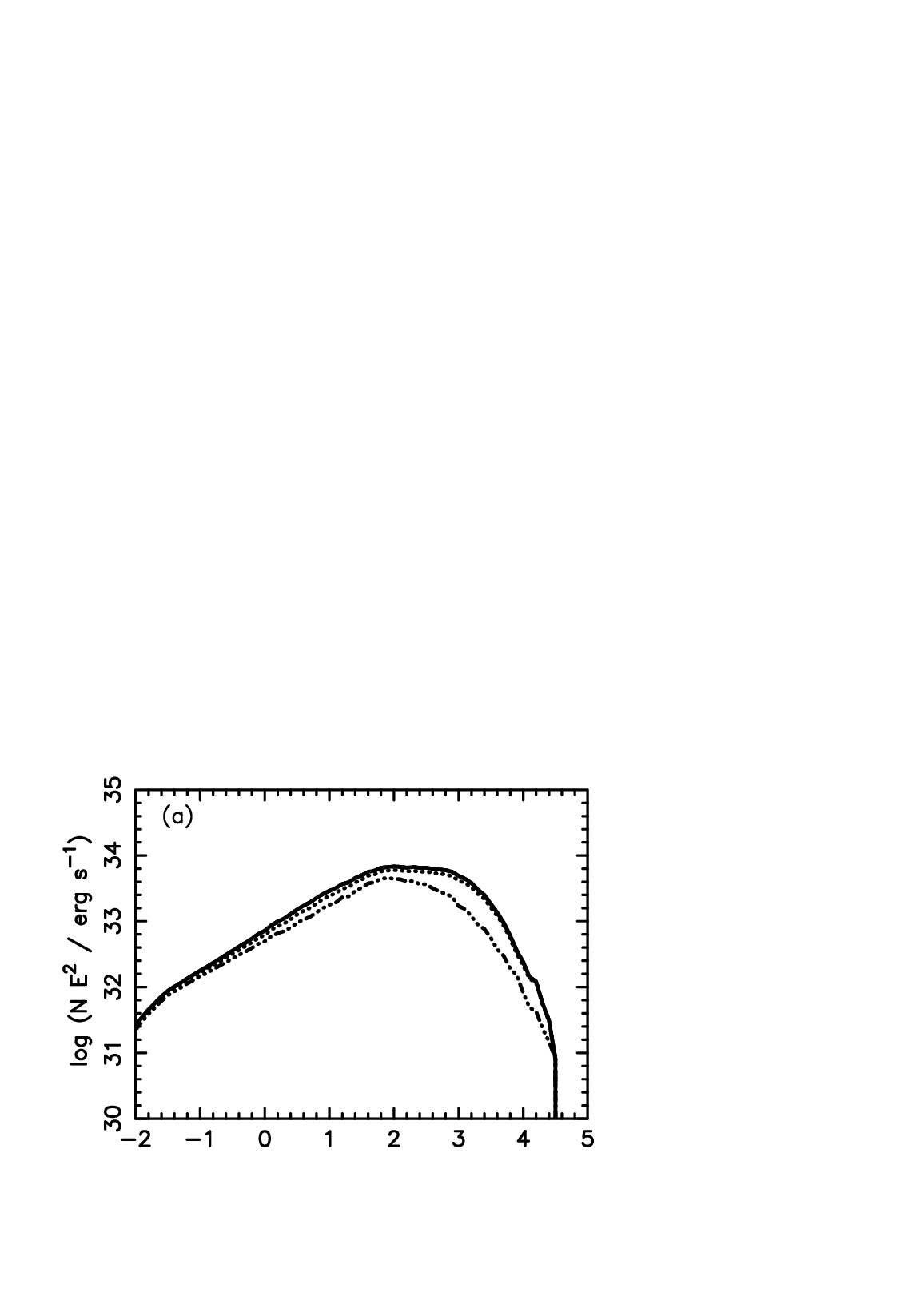}
\includegraphics{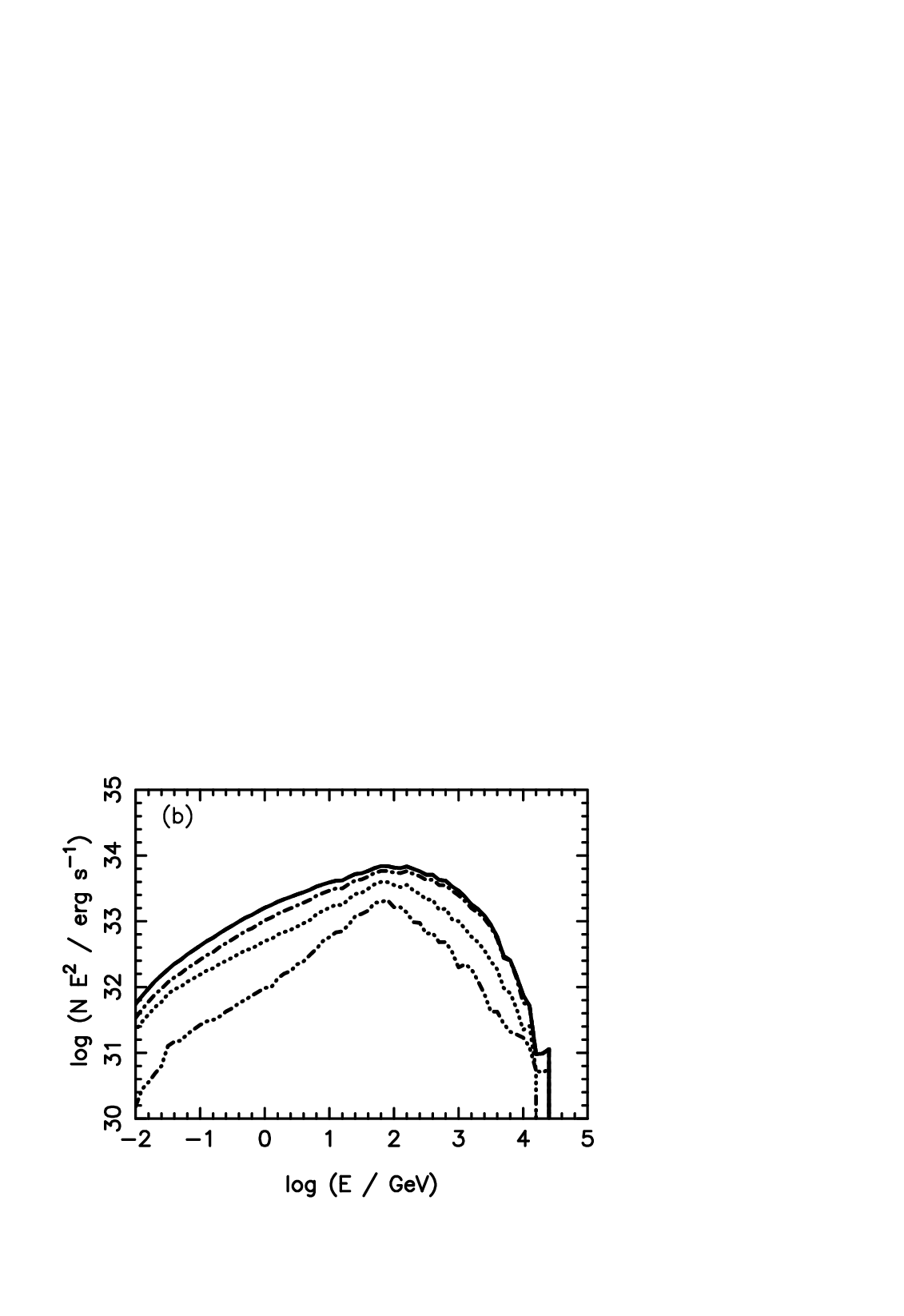}
\caption{The $\gamma$-ray spectra (multiplied by the energy squared) produced by leptons with the power law spectrum and spectral index $2.1$ (between 100 GeV to 30 TeV) within the central region of the GC with the radius R$<$1 pc (triple-dot curve), $<$3.3 pc (dotted), $<$10 pc (dot-dashed), $<$33 pc (dashed), and 100 pc (solid). The magnetic field inside the GC region is fixed on $B_{\rm GC} = 10^{-5}$ G (a) and $10^{-6}$ G (b). The specific GC has the core radius $R_{\rm c} = 0.5$ pc and the half mass radius $R_{\rm h} = 4$ pc.
The GC with the luminosity $L_{\rm GC} = 10^5$ M$_\odot$ is considered.}
\label{fig5}
\end{figure}

Although globular clusters are typically at large distances, the regions on the sky from which $\gamma$-rays can arrive to the observer at the Earth can be quite large, e.g. at the typical distance of 5 kpc and the half mass radius 5 pc the GC has angular dimension of $\sim 3'$. Therefore, it is important to find out whether the $\gamma$-rays produced by leptons diffusing outside the central region of GC can create a point like or an extended source for the Cherenkov telescopes.
In order to check this, we perform calculations of the $\gamma$-ray spectra as a function of distance from the center of the GC in case of their production by leptons with specific injection spectrum. The $\gamma$-ray spectra produced within the central volume with the radius 1 pc, 3.3 pc, 10 pc, 33 pc , and 100 pc are shown in Fig.~\ref{fig5}. They were calculated for  two different values of the magnetic field inside the GC, $B_{\rm GC} = 10^{-5}$ G (a) and $10^{-6}$ G (b).
It is clear that in the case of the magnetic field equal to $10^{-5}$ G, most of the $\gamma$-rays are already produced within $\sim 3$ pc from the center of the GC. The spectrum produced within $\sim$1 pc is only a factor of 2 lower than within $\sim$3 pc at energies above a few hundred GeV. In the case of the GC with the magnetic field equal to $10^{-6}$ G,
the $\gamma$-ray emission is more spreaded-out. However, still most of the emission ($\sim$50$\%$) comes from the central region with the radius of $\sim 3$ pc. Therefore,
we conclude that the extent of the $\gamma$-ray source, produced by leptons injected within the GCs, is relatively small provided that the magnetic fields inside the GC
cores are of the order of $10^{-6}-10^{-5}$ G. For typical distance to the GC (of the order of 5 kpc), the $\gamma$-ray source
with dimension of 3-5 pc should have an angular size of $\sim$2'-3', i.e. it should be a point-like source for the present Cherenkov telescopes that have the typical angular resolution of this order.
Since the extent of the $\gamma$-ray source is not very sensitive to the considered value of the magnetic field strength inside the GC, the more complicated model for the magnetic field structure inside the GC (e.g. dependence of the magnetic field on the distance from the center of GC) should not give very different results.

\section{Expected gamma-ray fluxes from specific objects}

As an example, we perform calculations of the $\gamma$-ray fluxes expected on the Earth from 4 globular clusters (selected from the catalog of globular clusters, Harris~1996). Inside these clusters many millisecond pulsars have been already detected and a large number of millisecond pulsars is expected (see Camilo \& Rasio 2005; Tavani~1993): 47 Tuc (22 MSP detected and $\sim 260$ MSP expected), Ter 5 (23, and $\sim 230$ MSP), M15 (8 and $\sim 117$ MSP), and M13 (5 and $\sim 102$ MSP). Moreover, these clusters have different declinations.
M15 and M13 can be observed at small zenith angles by the northern sky telescopes (MAGIC, and VERITAS), and Ter 5 and 47 Tuc can be observed by the southern telescopes (HESS and CANGAROO). M15 and Ter 5 can be observed simultaneously by the northern and southern sky  telescopes at small or large zenith angles what allows to cover the $\gamma$-ray energy range between $\sim$100 GeV and a few tens TeV. This is important if one wants to extract possible variable contribution to the $\gamma$-ray flux from e.g., X-ray binaries or intermediate mass black holes also expected inside the GCs. 

For every considered here cluster, we calculate the stellar radiation field as a function of distance applying their known parameters, i.e., dimension and total luminosity.
In all simulations we consider the propagation of leptons inside the inner region around the center of GC with the radius $R_{\rm max} = 10$ pc. As we have shown above, most of the $\gamma$-rays are produced in a relatively small region, provided that the magnetic field strength inside the cluster is larger than $B_{\rm GC} = 10^{-6}$ G. 
The $\gamma$-ray spectra are calculated for all these 4 clusters by assuming 
that the power injected in relativistic leptons is equal to $L_{\rm e^\pm} = 1.2\times 10^{35}$ erg s$^{-1}$. This power might be supplied by e.g.,
$N_{\rm p} = 100$ MSPs with the characteristic periods of 4 ms and surface magnetic fields of $10^9$ G, and assuming the efficiency of energy conversion from the pulsar winds into the relativistic leptons equal to $\eta = 0.01$ (see Eq.~\ref{eq9}). Note that 
the periods of MSPs detected in 47 Tuc are limited to a relatively narrow range around the average value of $\sim$4 ms (see Fig. 3 in Camilo \& Rasio 2005).
We assume that leptons are accelerated at the pulsar wind shocks with the power-law spectra ($\propto E^{-\alpha}$) and spectral indices in the range $\alpha = 2.1-3.0$, between $E_{\rm min} = 1$ GeV or $100$ GeV and $E_{\rm max} = 3$ TeV or 30 TeV. The applied energy range for spectrum of leptons has been motivated in Sect.~4.2.
The results of calculations for specific GCs mentioned above are discussed in a more detail below.

%
\begin{table*}
\caption{The upper limits on the parameter $N_{\rm p}\cdot \eta$, describing the injection of 
leptons into the globular cluster, 
based on the sensitivities of different telescopes (G - GLAST; M - MAGIC; H - HESS; V - VERITAS). }             
\label{tab1}      
\begin{tabular}{c c c c c c c c c c c c}     
\hline\hline       
Model          & 47 Tuc  & 47 Tuc   & Ter 5  & Ter 5  & Ter 5  & M 15 & M 15 & M 15 &  M 13  & M 13 & M 13    \\ 
$E_{\rm min}$, $\alpha$ & (G) & (H) &  (G)   & (M)    & (H, V) &  (G) & (M)  &(H, V)&  (G)   & (M)  & (V)  \\ 
\hline                    
100 GeV, 2.1  & 0.5    & 0.01       &  1.5   & 0.8    & 0.1    &  0.6 & 0.6  & 0.05  & 0.8 & 0.5 &  0.03 \\ 
\hline
100 GeV, 3.0  & 0.15   & 0.03       &  1     & 0.5    & 0.3    &  1.3 & 0.4  & 0.15  & 0.5 & 0.3 &  0.1 \\  
\hline
1 GeV, 2.1    & 0.1    & 0.03       &  2     & 2      & 0.15   &  0.5 & 1.5  & 0.1   & 0.3 & 0.8 &  0.05 \\  
\hline
1 GeV, 3.0    & 0.15   & 2          &  0.6   & 30     & 6      &  0.5 & 20   & 10    & 0.5 & 0.1 &  3 \\   
\hline                  
\hline
mono: 1 TeV   & 1  &  0.01  &  1  &  0.5 & 0.02  & 3 & 0.5 & 0.03  & 2  & 0.5  &  0.03 \\  
\hline
mono: 10 TeV  & 2  &  0.015 &  3 &  0.3  & 0.05  & 5 & 0.8 & 0.1   & 5  & 0.8  &  0.15 \\ 
\hline                  
\hline
\end{tabular}
\end{table*}
\subsection{47 Tuc  (NGC 104)}

47 Tuc is one of the closest globular clusters, at the distance of only 4.5 kpc. 
It belongs to the class of the most massive GCs in our Galaxy. 
It has a core radius $0.44$ arc min ($R_{\rm c} = 0.58$ pc), a half mass radius 
$2.79$ arc min ($R_{\rm h} = 3.67$ pc), and the total stellar luminosity $7.5\times 10^5$ L$_\odot$. 22 millisecond pulsars (with the periods in the range 2-6 ms) have been detected up to now within this cluster (e.g. Camilo et al.~2000). However, much larger number of MSPs, of the order of $\sim 200$ is predicted in 47 Tuc (see Tavani 1993, or more recently Ivanova, Fregeau \& Rasio 2004).

\begin{figure}
\vskip 9.5truecm
\includegraphics{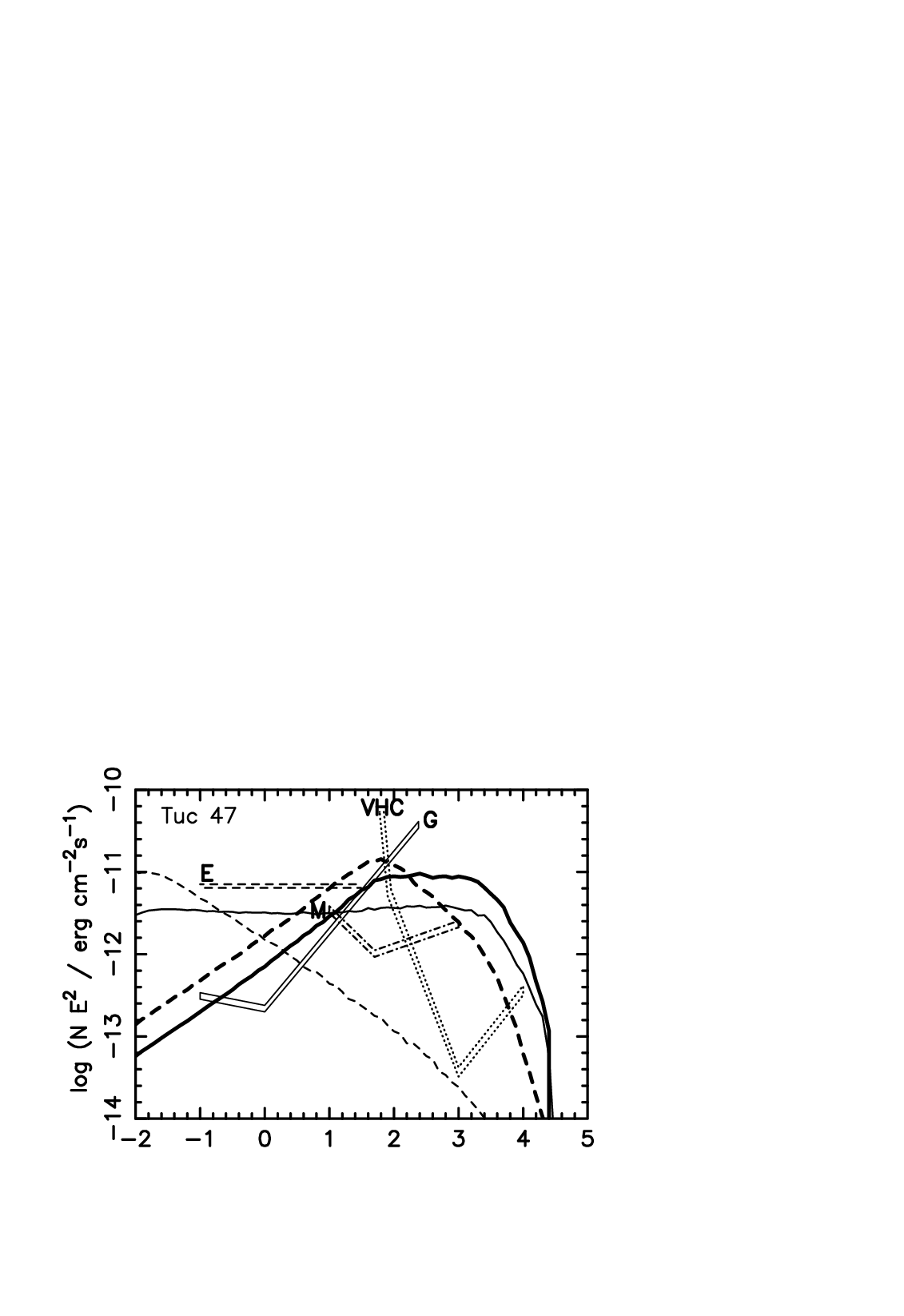}
\includegraphics{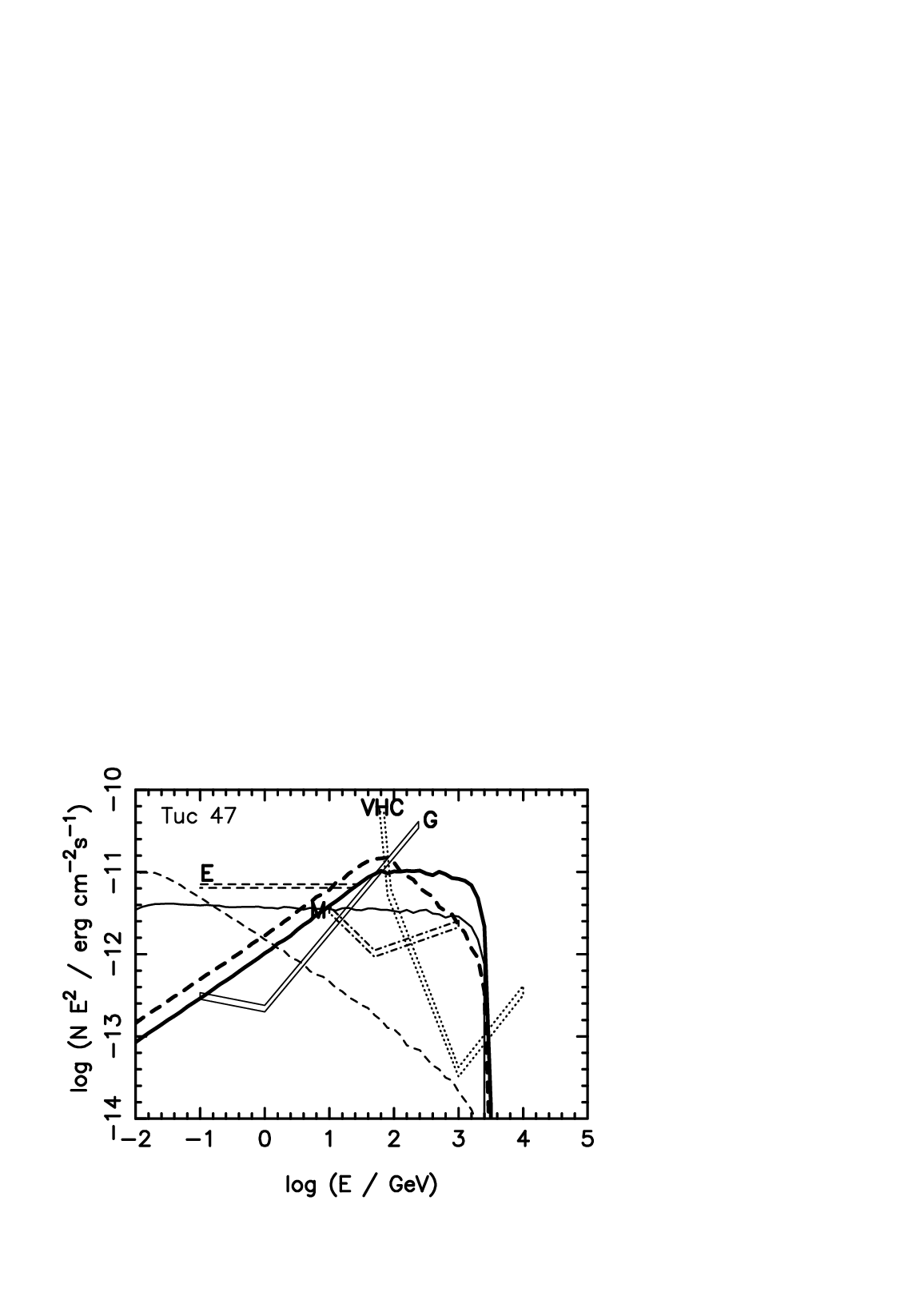}
\caption{The differential gamma-ray spectra (multiplied by the energy square, SED) expected
from 47 Tuc (NGC 104) for different parameters of the injected spectrum of leptons: power-law spectrum with index $\alpha = 2$ (solid curve), and $3$
(dashed). The spectrum has low energy cut-off at $E_{\rm min} = 1$ GeV (thin curves) or $100$ GeV (thick). $\gamma$-ray spectra produced by leptons with the upper energy cut-off at  $E_{\rm max} = 30$ TeV are shown on the upper figure and at $E_{\rm max} = 3$ TeV on the lower figure. The spectra are shown for the energy conversion efficiency from the pulsar winds to leptons equal to $\eta = 0.01$ and assuming the presence of $N_{\rm p} = 100$ millisecond pulsars inside the globular cluster ($N_{\rm p}\cdot \eta = 1$), with the parameters: the rotational periods equal to 4 ms and the surface magnetic fields equal to $10^9$ G. These pulsar parameters determine the power of the pulsar winds described by Eq.~\ref{eq9}. The available EGRET (E) upper limit for 47 Tuc (Michelson et al. 1994) is marked by a double dashed line.  
The sensitivities of different $\gamma$-ray telescopes ($5\sigma$ detection within 50 hrs,
Lorenz~2001) are marked by a double dotted curve (C - CANGAROO, H - HESS, and V - VERITAS), 
a double dot-dashed curve (M - MAGIC, and expected similar sensitivity of the HESS II in this energy range), and a double solid curve (G - GLAST).}
\label{fig6}
\end{figure}

The $\gamma$-ray spectra produced by leptons diffusing through the GC are shown in 
Fig.~\ref{fig6}ab, together with the sensitivities of the present and future $\gamma$-ray 
telescopes in the GeV and TeV energy range. For applied parameters of the acceleration 
scenario ($N_{\rm p} = 100$ MSP and $\eta = 0.01$), these $\gamma$-ray fluxes should be 
easily detected by the GLAST telescope provided that the spectrum of leptons extends below 100 GeV for both considered spectral indices. The Cherenkov telescopes can clearly detect $\gamma$-ray emission from this GC for the considered model parameters even if only $\sim$20 already observed MSPs are present inside 47 Tuc. The only  
exception is the case of a steep spectrum of leptons extending significantly below 100 GeV (see e.g., results for spectral index $\alpha = 3.0$).

In case of non-detection by the HESS telescope, the product $N_{\rm p}\cdot \eta$, which determines the flux level at $\gamma$-ray energies, 
should be $\le$0.01 for leptons with the spectrum extending between 100 GeV and
30 TeV, $\le$0.03 for the spectrum between 1 GeV and 30 TeV and spectral index equal to 2.1, and $\le$2 for the spectral index equal to 3. The upper limits on the product $N_{\rm p}\cdot \eta$, which can be obtained by the GLAST telescope, are usually not so restrictive (see details in Table 1). 
 
The available upper limits on the $\gamma$-ray flux in the MeV and GeV energy ranges from 
the direction of this cluster (mentioned in the introduction) do not put strong constraints 
on the number of the millisecond pulsars and efficiencies of lepton acceleration
(see e.g. Michelson et al.~1994, O'Flaherty et al.~1995, Manandhar et al.~1996).
For example, in order not to exceed the EGRET upper limit, the product 
$N_{\rm p}\cdot \eta$ can not be larger than few, i.e. still a few hundred MSP might be present inside 47 Tuc provided that energy conversion efficiency is of the order of $\eta = 0.01$.

\subsection{Ter 5}

\begin{figure}
\vskip 9.5truecm
\includegraphics{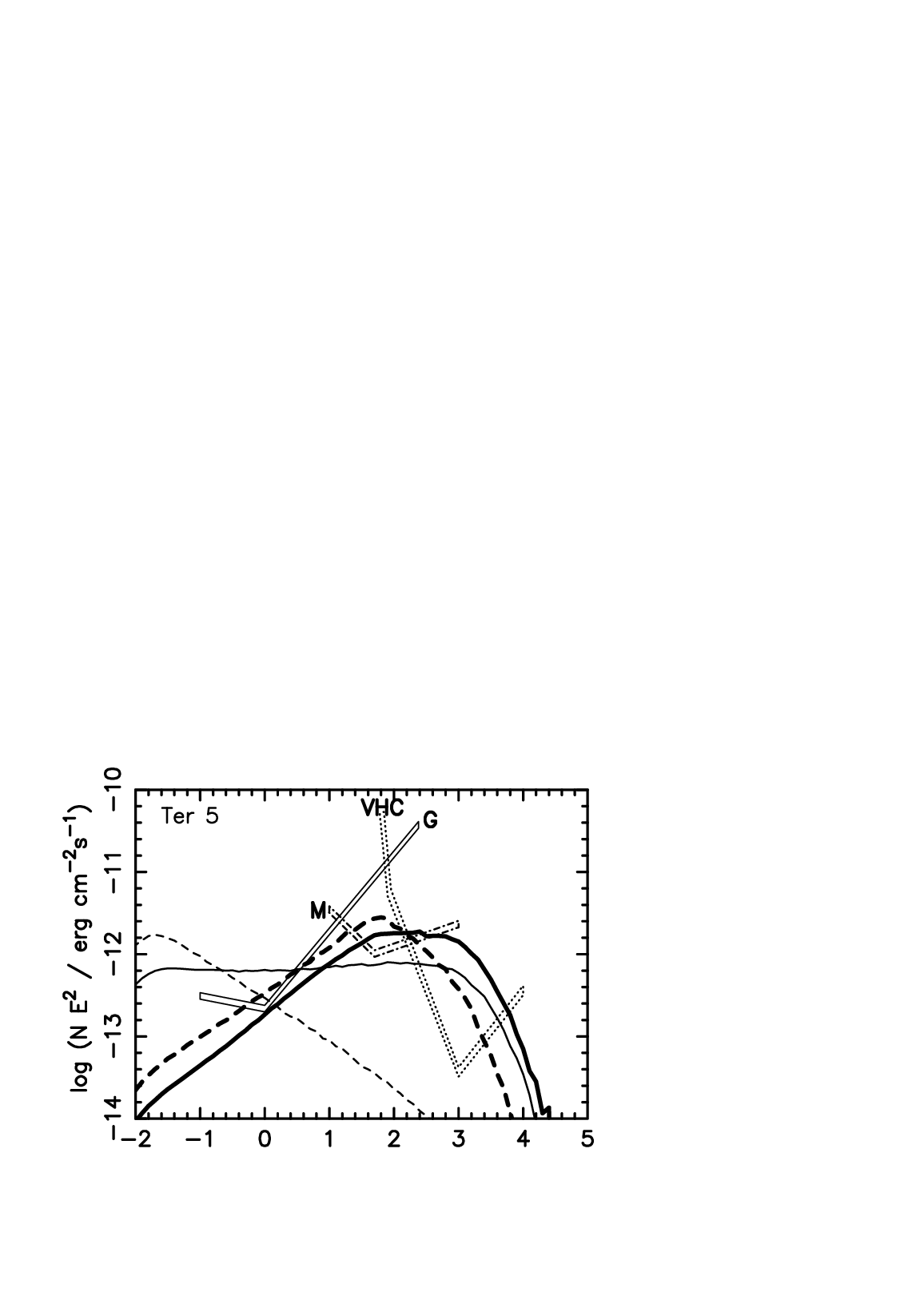}
\includegraphics{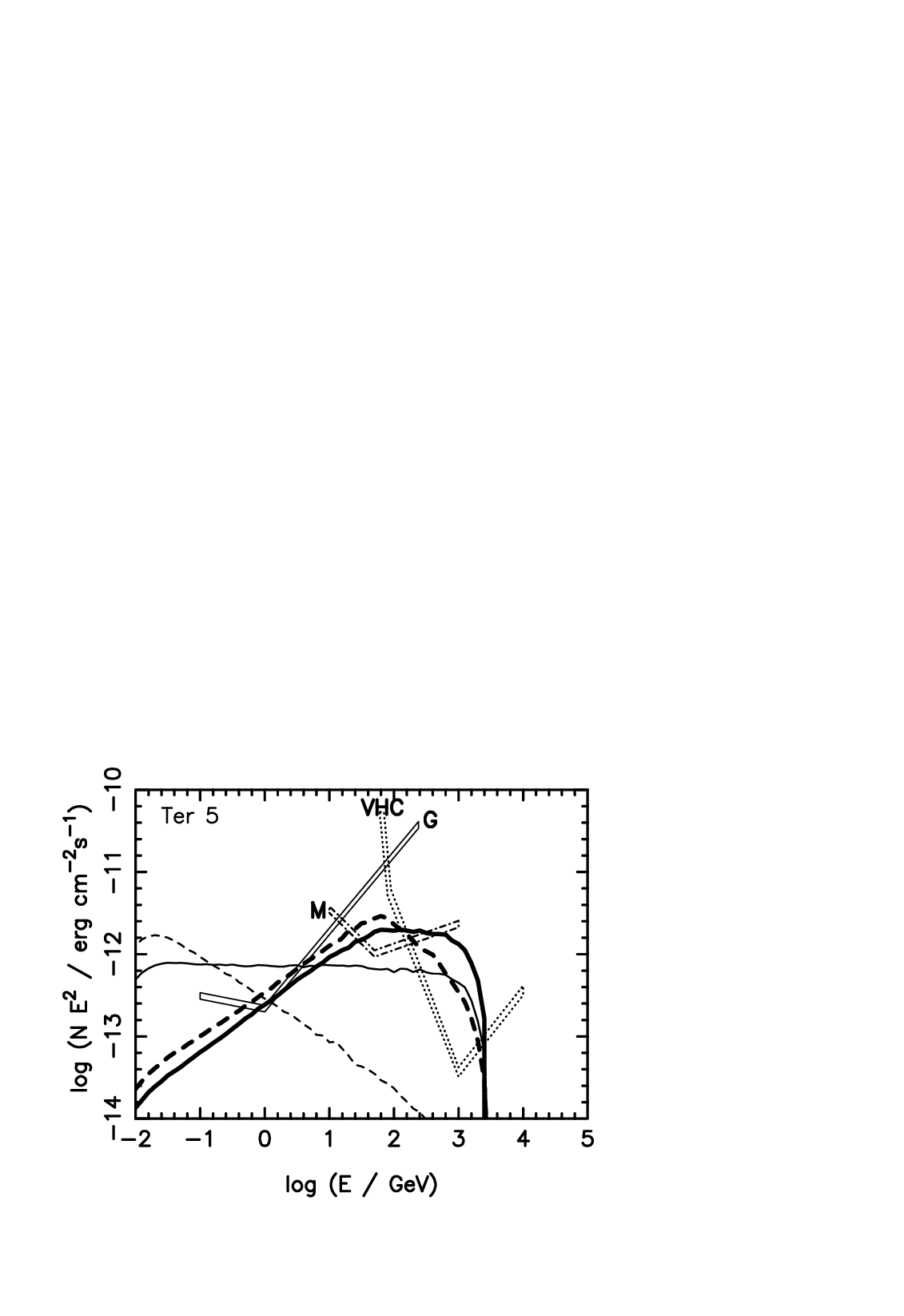}
\caption{As in Fig.~6 but for Ter 5.}
\label{fig7}
\end{figure}

Ter 5 has a core radius $0.18$ arc min ($R_{\rm c} = 0.54$ pc) and a half mass radius $0.83$ arc min ($R_{\rm h} = 2.5$ pc) at the distance of 10.4 kpc. Its total stellar luminosity is $1.5\times 10^5$ L$_\odot$. It contains 23 millisecond pulsars, the largest number detected so far within a single GC (Camilo et al.~2000).
Ter 5 has been observed by the EGRET telescope, but derived upper limit does not constrain significantly the population of the millisecond pulsars and the acceleration efficiencies of leptons (Michelson et al.~1994). 
The $\gamma$-ray spectra calculated for this GC and variety of considered models for lepton acceleration are shown in Fig.~\ref{fig7}. Due to the smaller luminosity and larger distance, these $\gamma$-ray spectra are on a significantly lower level than those previously estimated for 47 Tuc. They are close to the detection limit of the GLAST and MAGIC telescopes. However, Ter 5 has chance to be detected by the VERITAS, and HESS telescopes close to $\sim$1 TeV, i.e. at energies where the sensitivities of these telescopes are the largest. The eventual upper limits on the $\gamma$-ray flux from Ter 5 should constrain the acceleration efficiency of leptons inside this GC to the level of about one order of magnitude higher than in the case of 47 Tuc. Therefore, it seems that Ter 5 is not so promising GC from which GeV-TeV $\gamma$-ray emission might be observed in spite of the largest number of detected MSPs. Ter 5 is well situated at the large zenith angle for observations with the VERITAS telescopes and at low zenith angle for observations with the HESS telescopes what allows the coverage of a broadest possible energy range. Therefore, simultaneous observations at different zenith angles can provide useful information on possible variable $\gamma$-ray emission in the broad energy range. Note that discussed here model predicts steady $\gamma$-ray fluxes which should be easily distinguished from possible variable $\gamma$-ray fluxes expected in case of their origin e.g. in the low mass X-ray binaries.

\subsection{M15 (NGC 7078)}

M15 is at the distance of 9.4 kpc and belongs to the class of the core collapsed globular clusters. M15 is very compact with the core radius of 0.07 arc min ($R_{\rm c} = 0.19$ pc) and a half mass radius of 1.06 arc min ($R_{\rm h} = 2.91$ pc). 
The mass of the cluster is $5.4\times 10^5$ M$_{\odot}$ and the mass of the core inside $10^4$ AU is $3.4\times 10^3$ M$_{\odot}$. It is supposed  that the center of this globular cluster contains an $\sim 2000$ M$_\odot$ mass black hole (e.g., Gerssen et al.~2000). M 15 contains 7 known millisecond pulsars. However, the existence of about 100 millisecond pulsars is expected (Tavani~1993). 
Moreover, two interesting binary systems have been observed in M15. One of them, belonging to the LMXB class, X2127+119, is observed in X-rays with the power $\sim 10^{36}$ erg/s
(Beppo-Sax - Sidoli et al.~2000; Chandra - White et al.~2001). This system also shows huge outbursts with the peak power $6\times 10^{38}$ erg/s (Samle 2001). The second one is an ultra-compact binary with on orbital period of $\sim 17$ min, (Dieball et al.~2005). The Whipple group searched for WIMP annihilation signatures which might surround the core of M15. The upper limit on the rate of the  $\gamma$-ray photons, $\sim 0.2$ min$^{-1}$ can already weakly constrain some models of the WIMP mass and its distribution within M15 (Lebohec et al.~2003).

\begin{figure}
\vskip 9.5truecm
\includegraphics{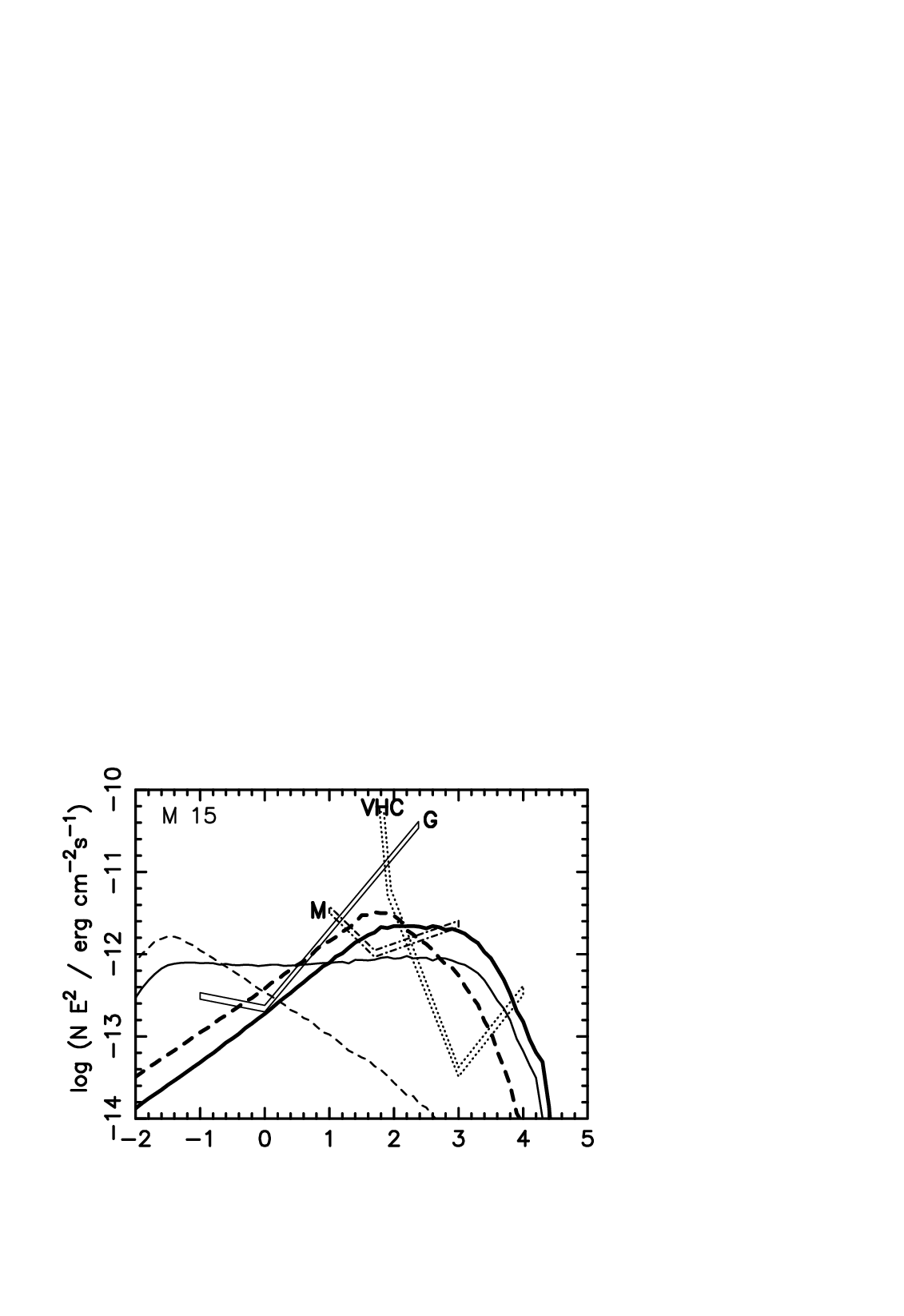}
\includegraphics{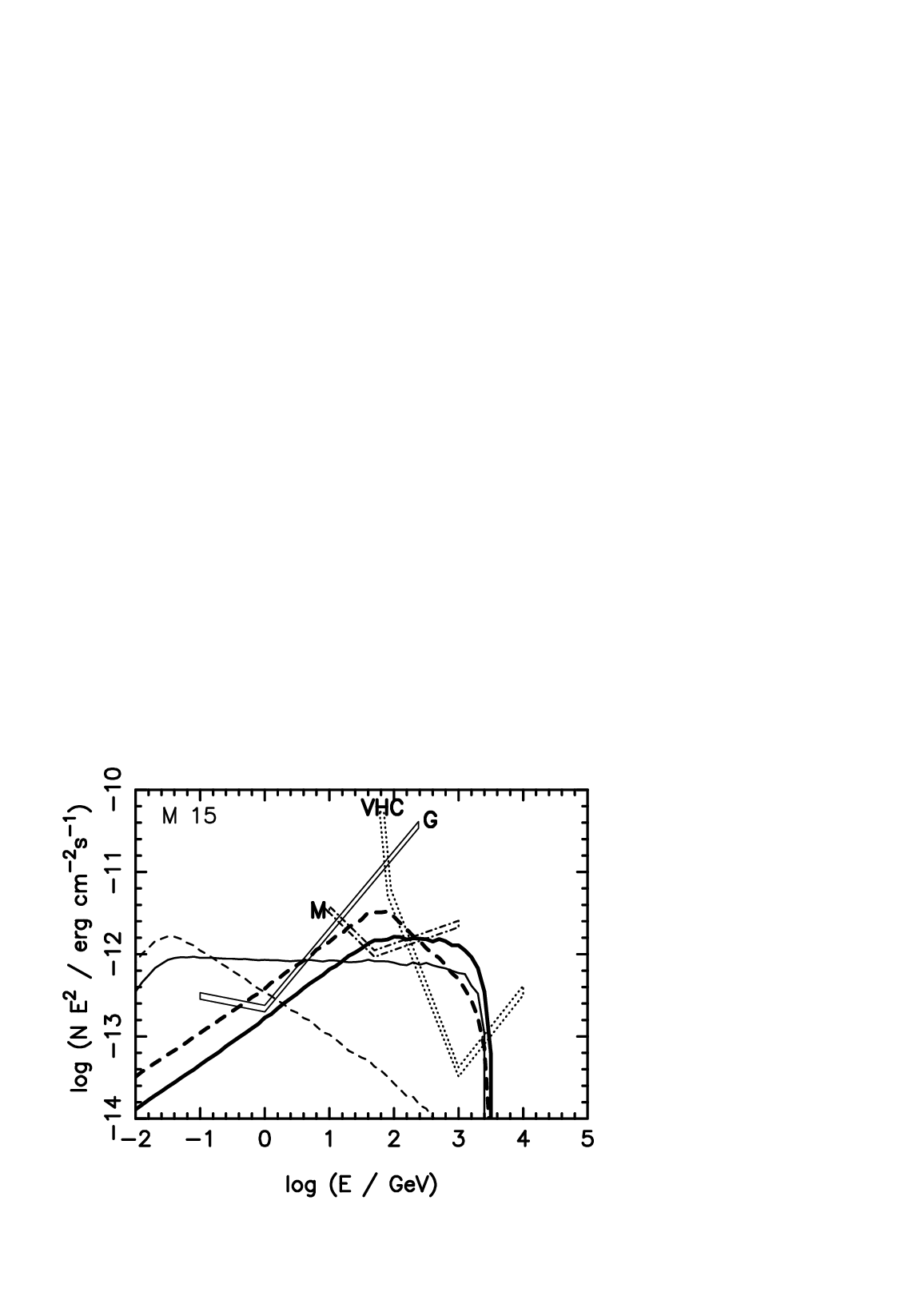}
\caption{As in Fig.~6 but for M 15.}
\label{fig8}
\end{figure}

We have calculated the $\gamma$-ray spectra expected from M15 in terms of the considered here injection scenarios of leptons.
Due to comparable distances and masses, they are very similar to these ones obtained from Ter 5. The strongest constraints on the number of MSPs and the energy conversion efficiency
($N_{\rm p}\cdot \eta$) are also introduced by the telescopes able to observe at TeV $\gamma$-ray energies, provided that leptons are accelerated to such energies (see, Table 1).
However, this GC is much better localized for the low zenith angle observations with the MAGIC and VERITAS telescopes. Therefore, it is more interesting when one considers the detection of the $\gamma$-ray signal in the case of the models which postulate acceleration of leptons with relatively steep spectra. M15 can be also observed at large zenith angle with the HESS telescope providing additional constraints on the spectrum of injected leptons and the $\gamma$-ray variability in the broad range. 
The GLAST observations in the GeV energy range can also put interesting limits on 
energy conversion efficiency (see Table 1) on the level below the present expectations from the theoretical modeling of pulsars. For example, if 100 MSPs are present, then energy conversion efficiency from the pulsar wind to relativistic particles can be limited by the GLAST sensitivity to $\eta < 0.005-0.013$ (Table 1). Note that the observations of the Crab Nebula limits $\eta$ to $\sim 0.1$. Therefore, this parameter can be more significantly constrained by the observations of MSPs in globular clusters.

Since, in principle, compact binary systems might also contribute to the high energy $\gamma$-ray emission from GCs, simultaneous observations in broad energy range will allow extraction of the non-variable part of $\gamma$-ray signal expected in terms of the present model.

\subsection{M13 (NGC 6295):}

M13 is at the distance of $\sim$7 kpc and  belongs to the class of normal globular clusters. It has a core radius of 0.78 arc min ($R_{\rm c} = 1.6$ pc) and a half mass radius of 1.49 arc min ($R_{\rm h} = 3.05$ pc). The total mass, $6\times 10^5$ M$_{\odot}$, is comparable to M15. Already 5 millisecond pulsars have been observed within this cluster (periods 2-10 ms) and the existence of 102 millisecond pulsars is predicted by Tavani~(1993). The upper limit on the $\gamma$-ray emission, $1.08\times 10^{-11}$ ph. cm$^{-2}$ s$^{-1}$ above 500 GeV, has been reported by the Whipple collaboration (Hall et al.~2003).

\begin{figure}
\vskip 9.5truecm
\includegraphics{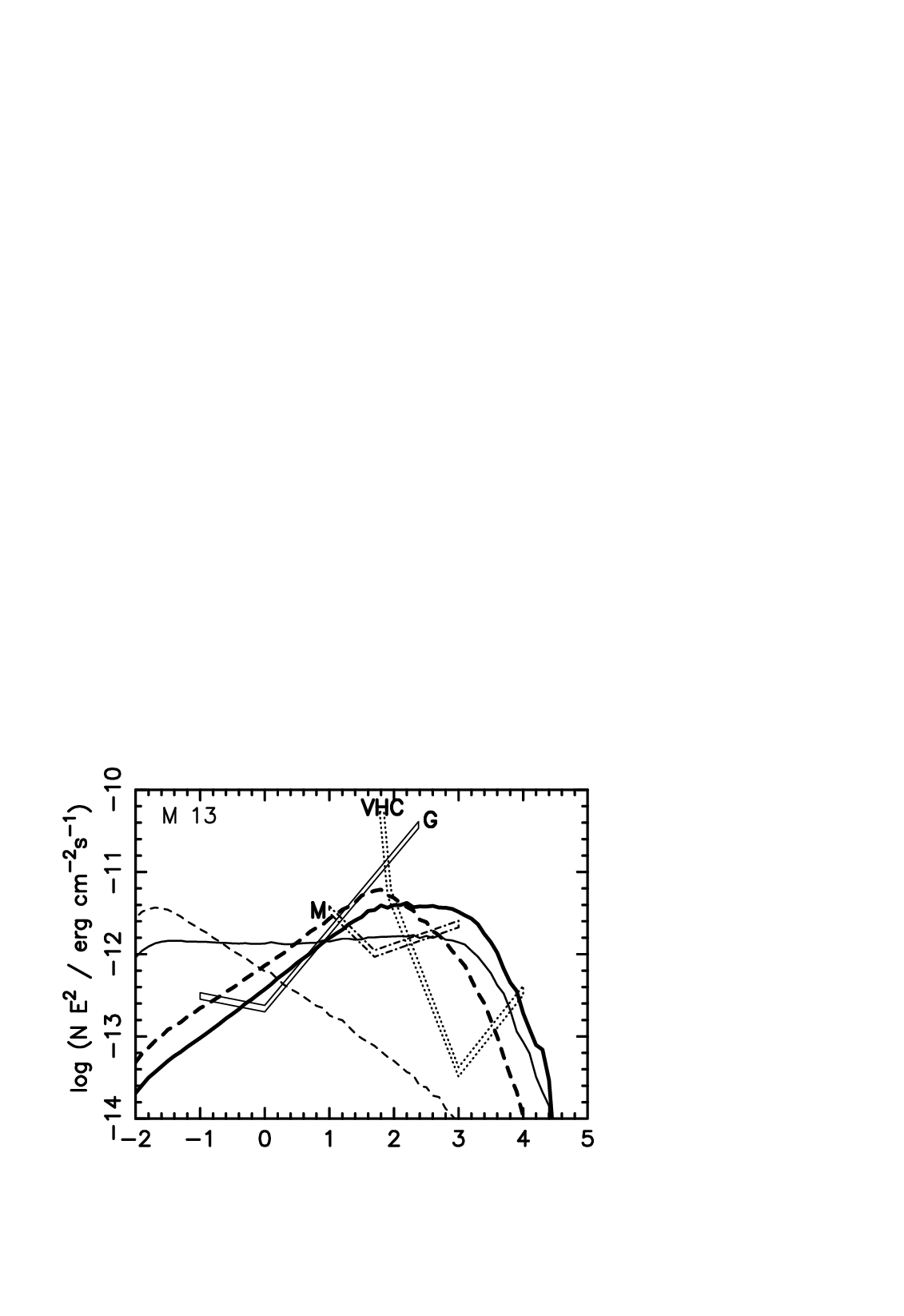}
\includegraphics{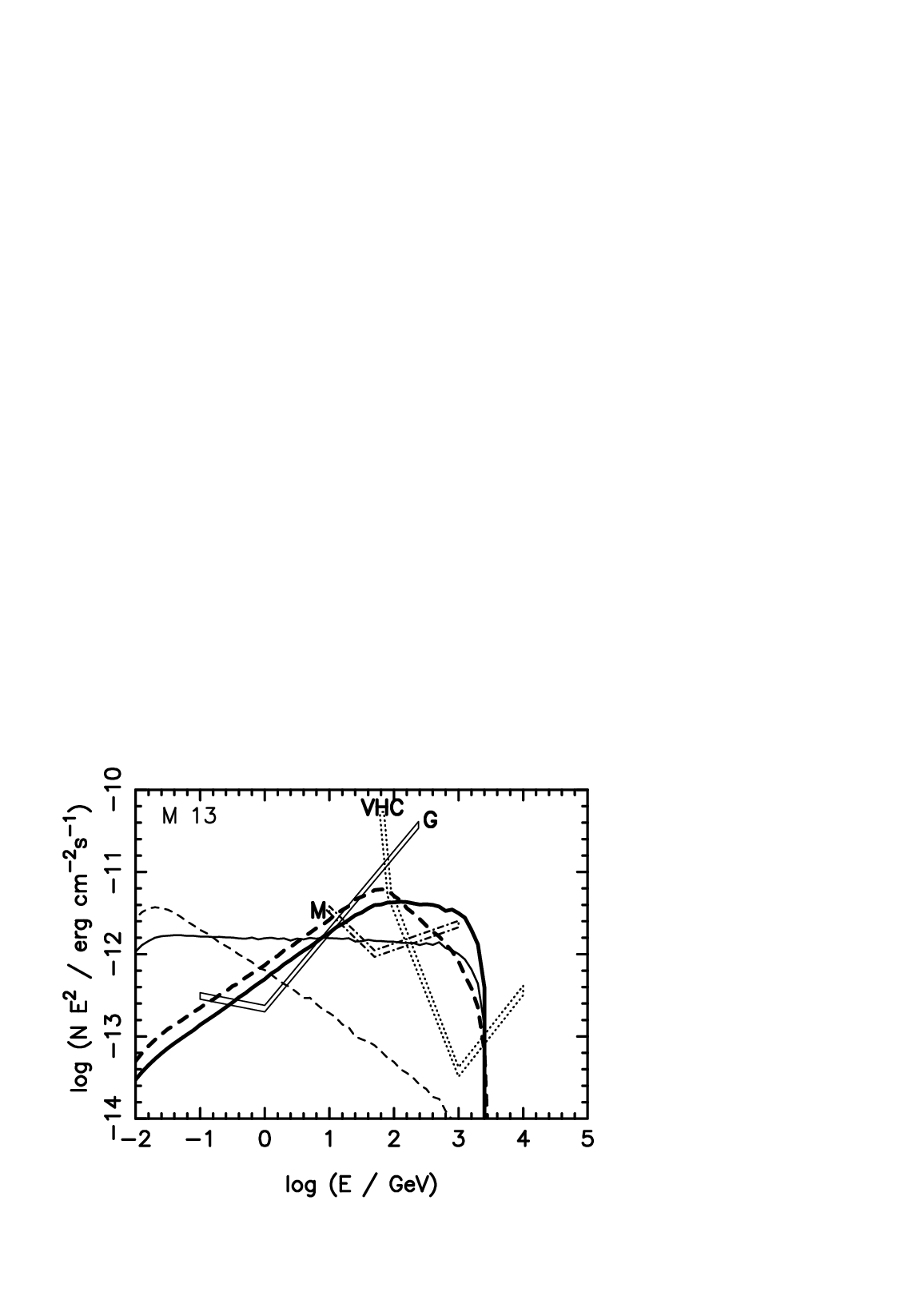}
\caption{As in Fig.~6 but for M 13.}
\label{fig9}
\end{figure}

The $\gamma$-ray fluxes from M13 calculated in terms of the present model are shown in Fig.~\ref{fig9}. These fluxes are a factor of $\sim$2 larger than in the case of M15, allowing to put stronger constraints on the number of MSPs and their energy conversion efficiency (see, Table 1). M13 can be observed at small zenith angles by the MAGIC and VERITAS telescopes. Hence, it is good candidate for detection in the sub-TeV energies in case of injection of leptons with relatively steep spectra or with energies lower than $\sim 1$ TeV. The GLAST and MAGIC telescopes should be able to constrain $N_{\rm p}\cdot \eta$ on the level of $<$0.1-1.0 (see Table 1). The constraints on $N_{\rm p}\cdot \eta$ introduced by the VERITAS observation can be an order of magnitude lower. Therefore, for 100 MSPs present in this GC, the energy conversion efficiency can be limited to $\sim 10^{-(3\div 4)}$. Note, that
the present upper limit put by the Whipple collaboration allows to constrain $N_{\rm p}\cdot \eta$ below $\sim$1.6, which for supposed number of 100 MSPs inside M13 allows to put the limit on $\eta\le 0.016$, which is about two orders of magnitude higher. 

\subsection{Leptons injected directly from the inner pulsar's magnetospheres}

In order to check whether the population of millisecond pulsars in specific GCs can
produce $\gamma$-ray fluxes potentially observed by the mentioned above telescopes, we 
also perform calculations of the $\gamma$-ray spectra assuming mono-energetic injection of leptons from the pulsar magnetospheres. These leptons can be accelerated inside the pulsar light cylinder radius as expected in the models for pulsed $\gamma$-ray emission, e.g, the extended polar gap model recently considered by Harding, Usov \& Muslinov~(2005).
The power in relativistic leptons, $L_{\rm e^\pm}$, which leave the inner magnetosphere through the light cylinder radius, is determined by the rotational energy loss rate of the millisecond pulsars, $L_{\rm p}$. 
Applying the recent model for acceleration of leptons inside the pulsar magnetopsheres, Muslinov \& Harding (1997) estimates this power on, $L_{\rm e^\pm} = 0.75\kappa (1-\kappa)L_{\rm p}\approx 0.1L_{\rm p}$, where $\kappa\approx 0.15$.  
The Lorentz factors of leptons escaping from the magnetopsheres of MSPs are expected to be limited by the curvature energy losses to values of the order of a few to several TeV (e.g. Bulik, Rudak \& Dysk 2000; Luo, Shibata \& Melrose 2000; Harding \& Muslinov~2002). In our calculations we apply the characteristic values for leptons injected into the GCs of the order of $1-10$ TeV.

\begin{figure}
\vskip 9.5truecm
\includegraphics{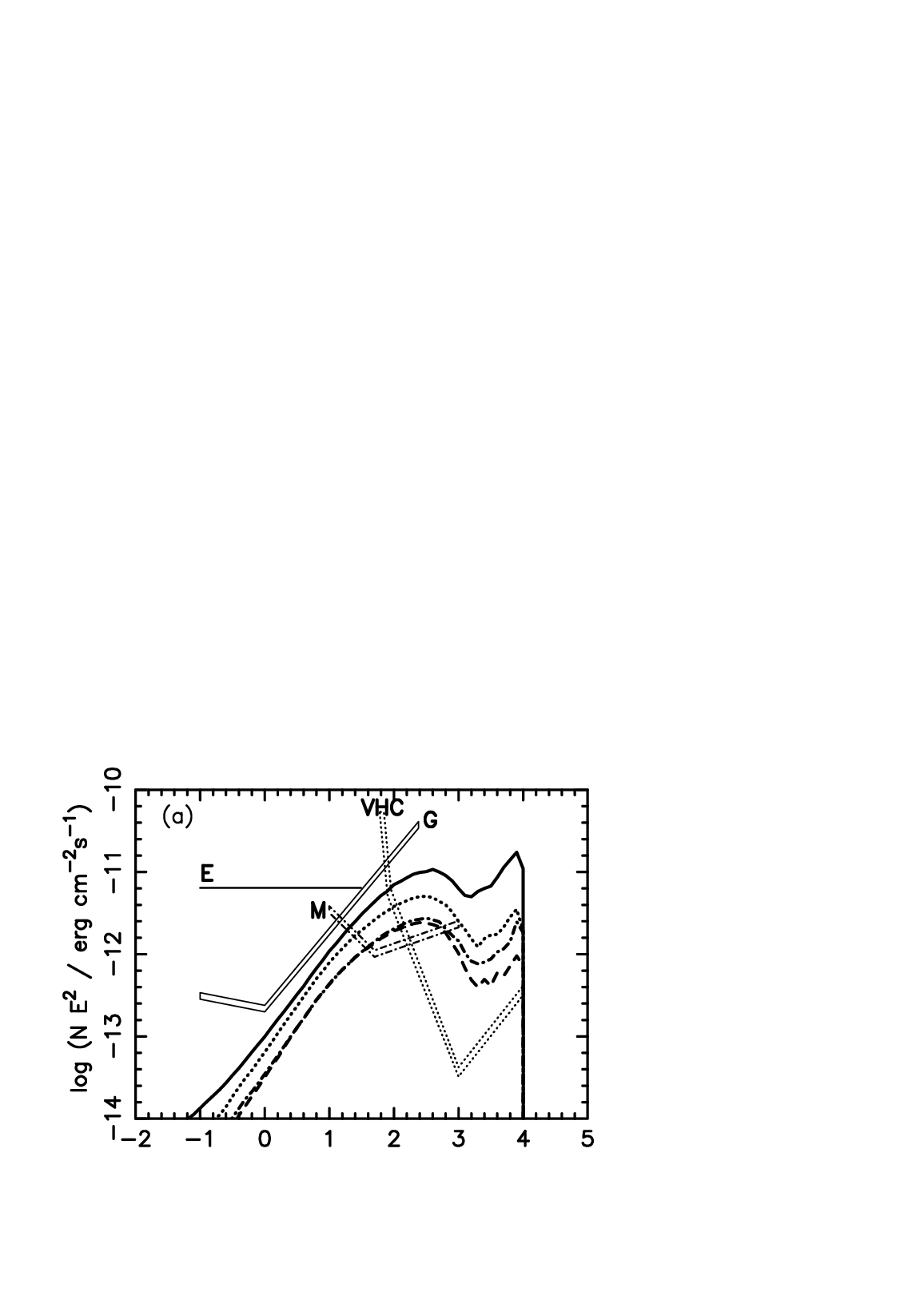}
\includegraphics{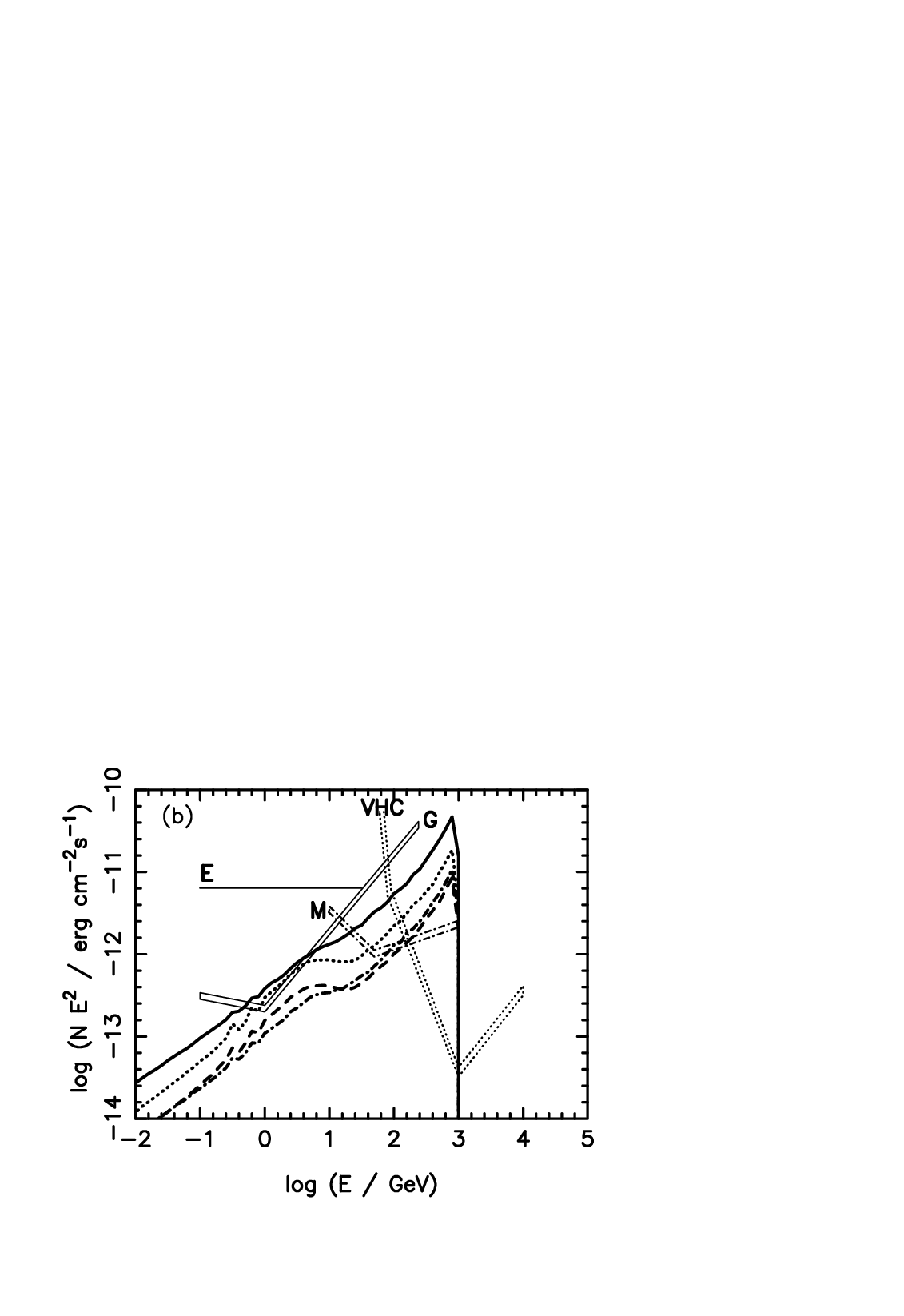}
\caption{The differential gamma-ray spectra (multiplied by the energy square, SED) expected
from 47 Tuc (solid curve), Ter 5 (dashed), M15 (dot-dashed), and M13 (dotted) in the case of mono-energetic leptons injected from the pulsar inner magnetosphere with energies 10 TeV (a) and 1 TeV (b). The $\gamma$-ray spectra are shown for the energy conversion efficiency from the pulsar winds to leptons equal to $\eta = 0.01$ assuming the presence of 100 MSP inside the GC. The sensitivities of different $\gamma$-ray telescopes and the EGRET upper limit  are marked by letters as in Fig.~\ref{fig6}.}
\label{fig10}
\end{figure}

For easier comparison with the above calculations, which assume acceleration of leptons at the pulsar wind shock with the power-law spectrum, we also assume the presence of 100 MSPs in specific GC and the efficiency of lepton acceleration equal to $\eta = 0.01$. The results of calculations for 47 Tuc, Ter 5, M15 , and M13, in the case of mono-energetic injection of leptons with energies 1 TeV and 10 TeV are shown in Fig.~\ref{fig10}. The $\gamma$-ray spectra expected from different GCs have similar shape but differ essentially in the  intensity. Note interesting double-peak structure evident in these spectra in the case of 10 TeV leptons.  They appear due to scattering of the MBR in the T regime (lower bump) and the scattering of stellar photons in the KN regime (narrower high energy peak). As in the case of the power-law spectra, the best candidates among GCs for positive detection with the Cherenkov telescopes are 47 Tuc (on the southern hemisphere) and M13 (on the northern hemisphere). 
Detection of the GeV $\gamma$-ray flux by the GLAST telescope in the case of mono-energetic injection of leptons seems to be less likely. However, if the acceleration efficiency of leptons is one order of magnitude larger than considered in this sample calculations, e.g. $\eta\sim 0.1$ as predicted by 
the  inner pulsar acceleration models (Harding, Usov \& Muslinov~2005), then also the GLAST telescope should detect the GeV $\gamma$-rays at least from 47 Tuc and M13.
In case of non-detection of any signal in the $\gamma$-ray energy range, the present and future telescopes can put strong limits on $N_{\rm p}\cdot \eta$ (see Table 1). 
These limits allow to constrain energy conversion efficiency from the pulsar to relativistic leptons below $\eta\sim 0.01-0.05$ (based on the GLAST observations in GeV energy range), $\eta\sim (3-8)\times 10^{-3}$ (the MAGIC observations at $\sim 100$ GeV), and $\eta\sim 10^{-(3\div 4)}$ (the HESS and VERITAS observations at $\sim 1$ TeV), provided that 100 MSPs are present in a specific GC (see Table 1).

\section{Discussion and Conclusions} 

Globular clusters are promising sites for constraining the acceleration processes of
leptons in the pulsar magnetopsheres and/or at the shock waves in the relativistic pulsar winds. Expected large injection rates of leptons by the whole population of millisecond pulsars inside GC have to interact with strong thermal radiation field, created by the huge concentration of stars, and with the microwave background radiation. Therefore, it is likely that GCs should be accompanied by a well localized, concentric, non-variable  $\gamma$-ray sources. Here, we consider two scenarios for injection of leptons and calculate expected 
$\gamma$-ray fluxes in the GeV-TeV energy ranges from a typical globular cluster. A few specific sources (47 Tuc, Ter 5, M15, M13), in which large population of millisecond pulsars is expected, are considered in detail. It is concluded that all these objects should be detected either by the GLAST (in the GeV energies) or Cherenkov telescopes (in the TeV energies) depending on the shape of the spectrum of injected leptons provided that: (1) leptons are accelerated up to $\sim$1 TeV; (2)
specific GCs contain at least several MSPs; and (3) the power conversion efficiency from the pulsar winds to relativistic leptons is comparable to that one observed in the case of the Crab Nebula ($\eta\sim 10\%$). It is expected, that in the case of acceleration of leptons in the inner pulsar magnetosphere's, $\eta$ is also of similar order, i.e. $\sim 3\%$ (Harding, Usov \& Muslinov~2002).

In the case of lack of positive detection of $\gamma$-rays from GCs, our calculations allow us to put strong constraints on the product of the power conversion efficiency times number of MSPs
in specific GC ($N_{\rm p}\cdot\eta$). In Table.~1 we show the upper limits on $N_{\rm p}\cdot\eta$ for specific GCs which may be introduced based on the sensitivities of the present Cherenkov telescopes and the future GLAST telescope for a few different models describing injection spectrum of leptons. If the acceleration of leptons occurs above $\sim 1$ TeV, then the most restrictive limits obtained by the HESS and VERITAS type Cherenkov telescope arrays on the power conversion efficiency of leptons are of the order of $\eta\sim 10^{-(3\div 4)}$, provided that $N_{\rm p}\sim 100$ MSPs are present in specific GC as expected in some estimations (e.g. Tavani 1993). The GLAST and MAGIC telescopes can provide also an order of magnitude more restrictive limits on $\eta$ than estimated from the observations of the Crab Nebula or expected in the case of isolated MSPs, provided that leptons are accelerated by $N_{\rm p}\sim 100$ MSPs to energies at least $\sim$1 TeV (Table.~1). These limits put strong constraints on the high energy processes occurring in the inner pulsar magnetopsheres and/or the acceleration mechanisms in the pulsar winds and the pulsar wind shocks. 

Note that considered here scenario differs from those ones which postulate production of
high energy $\gamma$-rays in the radiation processes occurring inside the inner pulsar magnetopsheres as proposed by e.g. Bulik, Rudak \& Dysk~(2000), Luo, Shibata \& Melrose (2000), and Harding, Usov \& Muslinov~(2005) or inside the interacting binary millisecond pulsars (Tavani 1993). In one of discussed here scenarios, in contrast to earlier model considered by Tavani~(1993), there is assumption that acceleration of 
leptons occurs at the part of the pulsar wind which is not limited by the wind of the companion star.  The second considered scenario with the mono-energetic injection of leptons, in contrast to the inner magnetosphere $\gamma$-ray production models mentioned above, takes into account the radiation processes from relativistic leptons which escaped from the inner pulsar magnetospheres and propagate in the whole volume of the GC.
$\gamma$-rays produced in terms of those earlier models can additionally contribute to the expected $\gamma$-ray flux mainly in the GeV energy range. Note that also other mechanisms of $\gamma$-ray production can operate inside the GCs. For example, it can occur that GCs contain microquasar type X-ray binaries (recently established as TeV $\gamma$-ray sources), intermediate  mass black holes, or related to them concentrations of dark matter particles. In fact, powerful X-ray binaries have been also observed inside some GCs, e.g. mentioned above X2127+119 inside M15 and some evidences of the presence of the intermediate mass black holes inside GCs have been also reported, e.g. in M15 (Gerssen et al.~2002). Therefore, the constraints on the number of MSPs and the power conversion efficiency from the pulsar winds to energetic leptons derived in this paper should be considered as the upper limits.

\section*{Acknowledgments}
This work is supported by the Polish MNiI grant No. 1P03D01028. 


\end{document}